\documentclass[oldversion]{aa}
\usepackage[dvips]{graphicx}
\usepackage{txfonts}
\usepackage[authoryear]{natbib}
\bibpunct{(}{)}{;}{a}{}{,}
\usepackage{bm}

\usepackage[multiple,perpage]{footmisc}

\begin{document}

\title{The line-of-sight proximity effect in individual quasar spectra%
  \thanks{Based on observations collected at the European
    Southern Observatory, Paranal, Chile (Programme 070.A-0425)} 
}

\author{Aldo Dall'Aglio \and Lutz Wisotzki \and G\'{a}bor Worseck}

\offprints{A. Dall'Aglio, \email{adaglio@aip.de}}

\institute{Astrophysikalisches Institut Potsdam, An der Sternwarte 16, D-14482 Potsdam, Germany}

\date{Received ... /  Accepted ...}

\abstract{We exploit a set of high signal-to-noise ($\sim 70$), low-resolution
  ($R\sim 800$) quasar spectra to search for the signature of the so-called
  proximity effect in the \ion{H}{i} Ly$\alpha$ forest. Our sample consists of 17
  bright quasars in the redshift range $2.7 < z < 4.1$. Analysing the spectra
  with the flux transmission technique, we detect the proximity effect in the
   sample at high significance. We use this to estimate the average
  intensity of the metagalactic UV background, assuming it to be constant over
  this redshift range.  We obtain a value of $J = (9 \pm 4) \times
  10^{-22}$ erg cm$^{-2}$ s$^{-1}$ Hz$^{-1}$ sr$^{-1}$, in good agreement with
  previous measurements at similar $z$. We then apply the same procedure to
  individual lines of sight, finding a significant deficit in the effective
  optical depth close to the emission redshift in every single object except
  one (which by a different line of evidence does nevertheless show a
  noticeable proximity effect). Thus, we clearly see the proximity effect as a
  universal phenomenon associated with individual quasars. Using extensive
  Monte-Carlo simulations to quantify the error budget, we assess the expected
  statistical scatter in the strength of the proximity effect due to shot noise
  (cosmic variance). The observed scatter is larger than the predicted one,
  so that additional sources of scatter are required. We rule out a dispersion
  of spectral slopes as a significant contributor. Possible effects
  are long time-scale variability of the quasars and/or gravitational
  clustering of Ly$\alpha$ forest lines. We speculate on the possibility of using the
  proximity effect as a tool to constrain individual quasar ages, finding
  that ages between $\sim 10^{6}$ and $\sim 10^{8}$ yrs might produce a 
  characteristic signature in the optical depth profile towards the QSO. 
  We identify one possible candidate for this effect in our sample.}


\titlerunning{The line-of-sight proximity effect in individual quasar spectra}
\authorrunning{A. Dall'Aglio et al.}

\maketitle

%---------------------------------------------------------------------------

\section{Introduction}
%------------------------------------------------------------------------------
%Figure 
%----------------------------------------------------------------------------
\begin{figure*}
\sidecaption
\includegraphics*[width=12cm]{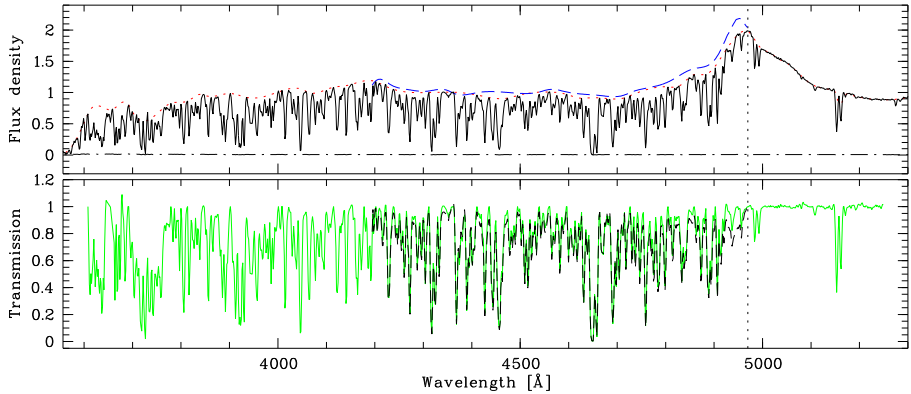}
\caption{Example of a quasar spectrum taken from the dataset (HE 0940$-$1050,
  $z=3.088$). In the upper panel the flux density in units of erg cm$^{-2}$ s$^{-1}$ \AA$^{-1}$
 as function of wavelength is
  presented with the uncorrected (dotted line), corrected (long-dashed line) continuum level  and the noise of the
  spectrum (dashed-dotted line). In the lower panel the fitted transmission is
  shown in gray while the corrected continuum is plotted in black (See Sect.~\ref{datared} and \ref{syserrors} for details). In both panels the vertical dotted line represent the emission redshift. }
\label{spec}
\end{figure*}
%------------------------------------------------------------------------------

The multitude of absorption lines seen in the spectra of high redshift
quasars gives important information about the state of matter in the universe,
tracing the physical conditions of the intergalactic medium (IGM) at
various epochs. It is commonly believed that for column densities up to
$N_{\mbox{\tiny\ion{H}{i}}} \approx 10^{17.5}\,\mathrm{cm^{-2}}$, the
absorbers are in photoionisation equilibrium with a metagalactic ultraviolet 
background field (UVB), composed of the integral over all sources 
of UV radiation (essentially, star-forming galaxies and quasars).
High-resolution spectra of the Lyman forest provide not only a rather
detailed statistical characterisation of the absorber properties
such as line number densities as well as temperature and density 
distributions \citep[e.g.,][]{kim01}, but also physical parameters
such as \ion{H}{i} and \ion{He}{ii} photoionisation rates 
\citep{rauch97,fardal98}, which directly relate to the intensity of the UVB, 
as described recently by \citet{bolton05} invoking hydrodynamical simulations.
Independently, the UVB has been successfully synthesised by combining the
observed quasar luminosity function and the UV emission from galaxies (although
the latter is still very uncertain) with the propagation of diffuse radiation 
\citep{haardt96}. 

In the vicinity of strong UV sources such as bright quasars, the \ion{H}{i}
photoionisation rate should locally increase, further reducing the density of
residual neutral hydrogen.  This enhances the transparency of the IGM to
\ion{H}{i} ionising radiation and should become observable as a weakening of
the Lyman forest absorption near such sources. Such an effect has first
been noted by \citet{carswell82} and was later baptised `Inverse'
\citep{murdoch86} or `Proximity Effect' \citet[][hereafter BDO88]{bajtlik88}.
Its prime application has been so far the possibility to derive an independent
estimate of the UVB intensity, by measuring the reduction of column
densities (against the global evolution of absorption line density increasing 
with redshift) and combining this with the QSO luminosity at the Lyman limit,
assumed to be known. The best constraints on the UVB using the proximity
effect stem from the combined analysis of large quasar samples 
\citep{cooke97,scott00,liske01}, yielding
mostly values consistent with the above quoted other methods. However, the
uncertainties are still substantial. Besides the problem that a limited number
of lines of sight always suffers from `cosmic variance', there may also be
systematic biases. In particular, if QSOs reside in intrinsically overdense 
environments then the signature of the proximity effect will appear weaker
than predicted \citep{loeb95,rollinde05}. Another uncertainty is the possibly limited
lifetime of quasars. On the other hand, the proximity effect may also be
used to derive constraints on this important, but largely unknown
astrophysical quantity \citep[e.g.,][]{pentericci02}.

In this paper we present an exploitation of new observational material in
terms of the proximity effect (Sect.~\ref{obs_set}). 
Rather than the traditional line counting we
use the more sensitive flux transmission statistic to search for proximity
effect signatures, augmented by extensive Monte-Carlo simulations to calibrate
the systematic and statistical errors (Sect.~\ref{lyaforest}). 
We deliver our results in Sections \ref{combined} and \ref{indiv_qsos}.
Firstly we briefly present an analysis of the combined sample of 17 QSO spectra
and derive an estimate of the UV background intensity. Secondly we demonstrate
that the effect is measurable on single sightlines \citep{williger94,lu96,savaglio97}, 
not only statistically in large samples \citep{bajtlik88, scott00, liske01}.

The effect can actually be systematically
detected in each single QSO of our sample (with one special case that is
discussed separately). We speculate about the possibilities to detect
signatures of finite quasar ages by virtue of the proximity effect.
 
Throughout this paper, we assume a flat $\Lambda-$Universe with
$\Omega_\mathrm{m}=0.3$ and $\Omega_\mathrm{\Lambda}=0.7$ and
$H_0=71\,\mathrm{km\,s^{-1}\,Mpc^{-1}}$.

%---------------------------------------------------------------------------
%Table 
%---------------------------------------------------------------------------
\begin{table*}[t]
\small\centering
\caption[]{Log of observations.}
\label{Tab1}
\begin{tabular}{lcccccclc}
\hline\hline\noalign{\smallskip}
QSO &   $V$ mag  &$z_\mathrm{em}$ &Exp. Time&Seeing & Airmass &Sky Condition$^\mathrm{a}$ & Obs. Date&Ref. Mag\\
&&&(s)&(arcsec)&&& \\
\noalign{\smallskip}\hline\noalign{\smallskip}
\object{CTQ 0247}        & 17.4 & 3.025 & 750  & 1.29 & 1.07 & CL, WI     & Dec. 8,  2002 & 4\\
\object{CTQ 1005}       & 18.4 & 3.205 & 1500 & 1.13 & 1.21 & TN         & Jan. 9,  2003 & 3\\
\object{CTQ 0460}        & 17.5 & 3.139 & 900  & 1.45 & 1.10 & PH         & Dec. 23, 2002 & 4\\
\object{H 0055$-$2659}   & 17.5 & 3.665 & 600  & 1.34 & 1.02 & CL, WI     & Dec. 8,  2002 & 4\\
\object{HE 0940$-$1050}  & 16.4 & 3.086 & 600  & 0.86 & 1.12 & TN, TK     & Nov. 26, 2002 & 3\\
\object{HE 2243$-$6031}  & 16.4 & 3.010 & 600  & 1.02 & 1.24 & CL, PH     & Nov. 9,  2002 & 4\\
\object{HE 2347$-$4342}  & 16.7 & 2.885 & 600  & 0.77 & 1.06 & PH         & Nov. 10, 2002 & 1\\
\object{PKS 2126$-$15}   & 17.0 & 3.285 & 600  & 0.97 & 1.20 & PH         & Oct. 29, 2002 & 4\\
\object{Q 0000$-$26}     & 18.0 & 4.098 & 600  & 1.31 & 1.03 & TN, CL     & Nov. 7,  2002 & 4\\
\object{Q 0002$-$422}    & 17.2 & 2.767 & 600  & 1.39 & 1.45 & TK, TN     & Oct. 14, 2002 & 3\\
\object{Q 0347$-$383}    & 17.7 & 3.220 & 800  & 1.45 & 1.15 & TN         & Jan. 9,  2003 & 1\\
\object{Q 0420$-$388}    & 16.9 & 3.120 & 600  & 1.37 & 1.06 & CL, WI     & Dec. 8,  2002 & 1\\
\object{Q 0913$+$0715}   & 17.8 & 2.787 & 800  & 1.45 & 1.18 & TN         & Jan. 9,  2003 & 2\\
\object{Q 1151$+$0651}   & 18.1 & 2.758 & 900  & 0.94 & 1.17 & TN, CL     & Jan. 25, 2003 & 2\\
\object{Q 1209$+$0919}   & 18.5 & 3.291 & 1500 & 0.78 & 1.80 & CL, TN, TK & Jan. 1,  2003 & 2\\
\object{Q 1223$+$1753}   & 18.1 & 2.945 & 900  & 0.66 & 1.35 & TN, CL     & Jan. 25, 2003 & 2\\
\object{Q 2139$-$4434}   & 17.7 & 3.214 & 800  & 0.85 & 1.33 & TN         & Apr. 30, 2003 & 3\\
\noalign{\smallskip}\hline\noalign{\smallskip}
\multicolumn{8}{l}{{}$^\mathrm{a}$ Legend: PH-Photometric, CL-Clear, TN-Thin cirrus, TK-Thick cirrus, WI-Windy.}
\end{tabular}
\begin{list}{}{}
\item[1:]\citet{worseck07}: PH conditions.
\item[2:]SDSS.
\item[3:]\citet{veron06}.
\item[4:]Slit loss corrected only.
\end{list}
\end{table*}
%------------------------------------------------------------------------------

\section{Data}\label{obs_set}

\subsection{Observations and data reduction}\label{datared}

Our data were obtained at the ESO-VLT UT4 (Yepun) in service mode between Oct
2002 and Apr 2003. We used FORS2 in long slit spectroscopy mode with the 600B
grism, covering the range 3315--6360\,\AA.  
A longitudinal atmospheric dispersion corrector (LADC) was used to account for 
differential refraction effects.
With a slit width of $1''$, the
resolution is $\sim 800$ (4\,\AA\ FWHM). Table~\ref{Tab1} summarises the
observations for the full sample. In total we observed 17 QSOs, always fully
covering the Ly Forest spectrum between the Ly$\alpha$ and Ly$\beta$ emission lines.

The spectra were reduced using IRAF standard procedures. Each exposure was
bias-corrected and flat-fielded, cosmic rays were marked using a
$\kappa$-$\sigma$ clipping algorithm, and the images were finally
sky-subtracted. The extracted spectra were calibrated in wavelength and flux
and corrected for vacuum and heliocentric shifts. 
The galactic extinction was taken into account assuming the $E(B-V)$ estimations by 
\citet{schlegel98}, together with the \citet{cardelli89} extinction curve assuming 
$R_\mathrm{V}=3.1$. 
The two exposures available for each target were coadded with inverse variance 
weights, yielding a typical signal-to-noise ratio (S/N) of $\sim 70$ in the Lyman forest.

\subsection{Quasar magnitudes}

An evaluation of the proximity effect relies on the accurate knowledge of quasar fluxes. Even though absolute spectrophotometry is compromised by intrinsic quasar variability, we reached reasonable accuracy in almost all spectra (See Tab.~\ref{Tab1} for details). We accounted for slit losses modelling a gaussian point spread function with FWHM given by the average seeing during the observations. Centering the 1 arcsec slit on the Gaussian centroid, we corrected for 
the flux falling outside the slit in all our spectra. In addition to our data, we used images obtained at ESO-VLT UT1 (Antu) in Nov 2004 under photometric conditions, covering 3 of our fields and SDSS information for 4 quasars yielding consistency to within $\delta V \mathrm{mag} \le 0.1$. For those spectra taken in clear to photometric conditions (6 in total), the corrected magnitudes matched the values from \citet{veron06} within $\delta V \mathrm{mag} \le 0.1$. The V magnitudes of the remaining objects, after slit loss correction, were systematically lower than the \citet{veron06} values by not more than 0.3 mag. Since those data were taken in relatively poor sky conditions, we adopted the \citet{veron06} values and associated larger uncertainties ($\delta V \mathrm{mag} \le 0.3$).

\subsection{Systemic quasar redshifts}

Our spectra cover a sufficient range in wavelength so that we could measure
the redshift of each quasar from more than one emission line.  All measured
redshifts are compiled in Table~\ref{Tab2}. In order to adopt a systemic
redshift we used low-ionisation lines whenever possible \citep{gaskell82,
tytler92}.  For objects where this was unfeasible, we used the redshift from
high-ionisation lines with a statistical correction, determined from the
average shift between \ion{Si}{ii}+\ion{O}{i} and \ion{Si}{iv}+\ion{O}{iv]}.
The agreement between \ion{Si}{ii}+\ion{O}{i} and \ion{C}{ii} estimates is
generally good, even though the second line is usually very weak and rather
broad. For Q~0000$-$26 only the asymmetric Ly$\alpha$ line with strong associated
absorption was covered by our spectrum; for this object we adopted the
redshift from \citet{Schneider91}. For Q~0347$-$383 our corrected value 
is in agreement with the estimation done by \citet{steidel90}. 

\subsection{Quasar continuum}

The analysis of absorption lines requires a normalisation to the QSO
continuum.  We explored two types of continuum estimates: (i) a global power
law ($f(\nu) \propto \nu^\alpha$), excluding emission and absorption regions,used 
to estimate the quasar flux at the Lyman limit; (ii) a
more local estimate that also includes the broad emission lines as
quasi-continuum. For this task we developed an automatic algorithm, following the 
work by \citet{young79} and \citet{carswell82}, which perform a cubic spline 
interpolation based on adaptive intervals along the spectrum with respect to the 
continuum slope. The points for the spline interpolation were chosen starting from a regular sampling of the spectrum with a binning that becomes finer whenever the slope of the computed continuum exceeds a given threshold. This is done in order to better reproduce the wings of emission lines.

In Sect.~\ref{syserrors} below we assess the expected errors (arising mainly from
line crowding) associated with this process.  Figure~\ref{spec} shows a sample
quasar spectrum together with the estimated local continuum and the resulting
transmission spectrum (see online material for the complete set of quasar spectra).
%---------------------------------------------------------------------------
%Table 
%------------------------------------------------------------------------------
\begin{table*}[t]
\small\centering
\caption{Redshift estimates with associated errors for the quasar sample resulting from different
  emission lines$^{\dagger}$.}
\label{Tab2}
\begin{tabular}{lcccccccc}
\hline\hline\noalign{\smallskip}
QSO& $z_{\mathrm{Ly}\alpha}$ &$z^\mathrm{a}_{\mbox{\tiny\ion{Si}{ii}}+\mbox{\tiny\ion{O}{i}}}$&$z_{\mbox{\tiny\ion{C}{ii}}}$ &$z_{\mbox{\tiny\ion{Si}{iv}}+\mbox{\tiny\ion{O}{iv]}}}$&$z_{\mbox{\tiny\ion{C}{iv}}}$& $\Delta z$&$z_\mathrm{Lit}$  & Reference\\
& $\sigma_z=0.003$ &$\sigma_z=0.003$&$\sigma_z=0.005$ &$\sigma_z=0.003$&$\sigma_z=0.003$&$z_{\mbox{\tiny\ion{Si}{ii}}+\mbox{\tiny\ion{O}{i}}} - z_{\mbox{\tiny\ion{Si}{iv}}+\mbox{\tiny\ion{O}{iv]}}}$&\\
\noalign{\smallskip}\hline\noalign{\smallskip}
CTQ 0247          & 3.008 & 3.025               & 3.021 & 3.016 & 3.008  & 0.009 & 3.020  & 1\\
CTQ 1005          & 3.201 & 3.205               & 3.210 & 3.196 &   -    & 0.009 & 3.210  & 1\\
CTQ 0460          & 3.134 & 3.139               & 3.135 & 3.128 &   -    & 0.011 & 3.130  & 1\\
H 0055$-$2659     & 3.650 & 3.665               & 3.659 &  -    &   -    &   -   & 3.655  & 2\\
HE 0940$-$1050    & 3.081 & 3.086               & 3.088 & 3.074 & 3.059  &   -   & 3.068  & 3\\
HE 2243$-$6031    & 3.004 & 3.010               & 3.008 & 3.005 & 3.004  & 0.005 & 3.010  & 4\\
HE 2347$-$4342    & 2.877 & 2.885               & 2.886 & 2.870 & 2.861  & 0.015 & 2.885  & 5\\
PKS 2126$-$15     & 3.279 & 3.285               & 3.286 & 3.270 &   -    & 0.015 & 3.267  & 3\\
Q 0000$-$26       & 4.100 &   -                 &   -   &   -   &   -    &   -   & 4.098  & 6\\
Q 0002$-$422      & 2.765 & 2.767               & 2.765 & 2.757 & 2.756  & 0.010 & 2.767  & 3\\
Q 0347$-$383      & 3.213 & 3.220$^\mathrm{b}$  &   -   & 3.209 &   -    &   -   & 3.222  & 12\\
Q 0420$-$388      & 3.117 & 3.120               & 3.120 & 3.109 &   -    & 0.011 & 3.110  & 2\\
Q 0913$+$0715     & 2.784 & 2.787               & 2.786 & 2.779 & 2.767  & 0.008 & 2.785  & 7\\
Q 1151$+$0651     & 2.755 & 2.758               &   -   & 2.752 & 2.754  & 0.006 & 2.762  & 8\\
Q 1209$+$0919     & 3.292 & 3.291               & 3.289 & 3.278 &   -    & 0.013 & 3.291  & 9\\
Q 1223$+$1753     & 2.935 & 2.945               & 2.944 & 2.936 & 2.930  & 0.009 & 2.936  & 10\\
Q 2139$-$4434     & 3.211 & 3.214               & 3.210 & 3.197 &   -    & 0.017 & 3.230  & 11\\
\noalign{\smallskip}\hline%\noalign{\smallskip}
\end{tabular}
\begin{list}{}{}
\item[$\dagger$:] Wavelength used to estimate the redshifts are: $\mathrm{Ly}\alpha = 1215.67$\AA, $\tiny\ion{Si}{ii}+\tiny\ion{O}{i}=1305.77$\AA, $\tiny\ion{C}{ii}=1335.30$\AA, $\tiny\ion{Si}{iv}+\tiny\ion{O}{iv]}=1396.76$\AA, $\tiny\ion{C}{iv}=1549.06$\AA \citep{morton03}
\item[a:] Taken as systemic redshift.
\item[b:] These redshifts were computed from the average shift between the redshift measurements of the \ion{Si}{ii}+\ion{O}{i} and \ion{Si}{iv}+\ion{O}{iv]} emission lines. The average redshift shift is about 0.011.
\item[Ref:] (1) \citet{Lopez01}, (2) \citet{Osmer94}, (3) \citet{rollinde05}, (4) \citet{Lopez02}, (5) \citet{reimers97}, (6) \citet{Schneider91}, (7) \citet{Pettini97}, (8) \citet{veron06}, (9) \citet{Storrie00}, (10) \citet{Hewett95}, (11) \citet{Hawkins93}, (12) \citet{steidel90}.
\end{list}
\end{table*}
%------------------------------------------------------------------------------

\section{Analysis}\label{lyaforest}

\subsection{The flux transmission technique}

The distribution of absorption lines along a line of sight (LOS) towards a quasar is
usually expressed as a function of redshift $z$, column density $N_{\mbox{\tiny\ion{H}{i}}}$, and Doppler
parameter $b$ in the form
$\mathrm{d^3}n/\left(\mathrm{d}z\mathrm{d}N_{\mbox{\tiny\ion{H}{i}}}\mathrm{d}b\right)=\eta(z,N_{\mbox{\tiny\ion{H}{i}}},b)$.
Due to our limited spectral resolution we could not perform single absorber
analysis. We followed instead the approach proposed by \citet{Zuo93b} and
\citet{liske01} to link the line number density to the evolution of the
effective optical depth. The resulting evolution depends on redshift as
\begin{equation}
\tau_\mathrm{eff}= B(1+z)^{\gamma+1}\label{eq:tau}
\end{equation}
where $B$ and $\gamma$ are sensitive to the resolution and the detectable
column density range and the observable effective optical depth defined as the optical 
depth at the average transmission over a predefined wavelength interval: 
$e^{-\tau_\mathrm{eff}}=<e^{-\tau}>$.

In order to account for local fluctuations of the ionising radiation field, we
follow the approach by BDO88 which assumes intervening absorbers
to be in photoionisation equilibrium with the local ionising field;
furthermore an empty space, and no flux attenuation except geometric dilution.
The modification introduced in the optical depth then becomes
\begin{equation}
\tau_\mathrm{eff}=B(1+z)^{\gamma+1}(1+\omega)^{1-\beta}
\end{equation}
where $\omega$ is the ratio between the photoionisation rates of the quasar
and the background and $\beta$ the slope in the column density distribution. 
Assuming a constant UVB in the range of redshifts of our
sample, an equal spectral energy distribution of QSOs and background at
$\nu>\nu_\mathrm{LL}$, and pure hydrogen absorbers that are isothermal, 
homogeneous, and randomly distributed along the LOS, we find 
\begin{equation}\label{34.6}
\omega=\frac{f_{\nu}(\lambda_\mathrm{LL}(1+z_{\mathrm{c}}) )}{4\pi J_{\nu}}\frac{1}{(1+z_{\mathrm{c}})}
        \left(\frac{d_{L}(z_{\mathrm{q}},0)}{d_{L}(z_{\mathrm{q}},z_{\mathrm{c}})}\right)^{2} 
\end{equation}
with $z_{\mathrm{c}}$ being the cloud redshift, $d_{L}(z_{\mathrm{q}},0)$ the
luminosity distance of the QSO as seen from the Earth, and
$d_{L}(z_{\mathrm{q}},z_{\mathrm{c}})$ as seen from the cloud. As clarified by
\citet{liske01}, $f_{\nu}(\lambda_\mathrm{LL}(1+z_{\mathrm{c}}))$ is the flux
at the Lyman limit which has to be weighted by a bandwidth correction. We
computed then the \emph{normalised optical depth} (also called $\xi$) which is
the deviation of the detected optical depth from the one expected in the Ly Forest 
\begin{equation}
\xi=\frac{\tau_\mathrm{eff}}{B(1+z)^{\gamma+1}}=(1+\omega)^{1-\beta}. 
\label{eq:xi}
\end{equation}

In order to quantify the reduction of $\xi$ close to a QSO we need to know the
parameters $B$ and $\gamma$ quantifying the redshift evolution of the
Lyman forest. These values are typically determined from high-resolution
spectroscopy. \citet{kim02} obtained $B=0.0032$ and $\gamma = 2.37$, which we
used as starting values to compute the normalised optical depths for all QSOs
over the full spectral range. This resulted in slightly too high average $\xi$
outside the proximity effect zones, where Eq.~\ref{eq:tau} should hold and
produce a mean $\xi$ of unity. We corrected this slight mismatch by adjusting
$B$  until we reached $\xi_{\omega \rightarrow 0} \sim 1$ for the Ly$\alpha$ forest region in
all spectra, excluding the proximity effect zone. This resulted in final
normalisation parameter of $B = 0.0041$. 
We will quantify the impact of different normalisation strategies when presenting 
the results for the UV background in Sect.~\ref{combined}.
Finally the slope of the column density distribution was set throughout the paper
to $\beta=1.5$ \citep[e.g.][]{hu95,kim02}, if not explicitly written.

In order to reveal the proximity effect, we now searched for a 
systematic departure of the normalised optical depths $\xi$ from unity 
for large values of $\omega$. The result for our combined sample is shown 
in Fig.~\ref{pe_res}, while Fig.~\ref{PE_prof_sample} displays the run
of $\xi$ versus $\omega$ for each individual QSO line of sight. Before we 
discuss these results we want to briefly describe our approach to quantify
the statistical and systematic errors.

\subsection{Error estimates from synthetic spectra}\label{errors}

\subsubsection{Method}\label{method}

Realistic models of the Ly forest, as already developed by
\citet{Zhang95}, invoke the baryonic component in CDM simulations to trace
the absorption line properties along sight lines towards QSOs. However, in a
first approximation, such distributions can be considered as random processes,
mathematically governed by Poisson statistics. Following this assumption
we performed extensive Monte-Carlo simulations to study the error budget,
in particular systematic errors arising from limited spectral resolution.
 
Each given simulated line of sight was populated with lines distributed
using $\mathrm{d}n / \mathrm{d}z \propto (1+z)^{\gamma}$ as line 
number density distribution leading to $ \tau_\mathrm{eff}= B(1+z)^{\gamma+1}$. 
The algorithm continues to add absorption features until the effective optical depth 
reaches the expected value of $\tau_\mathrm{eff}$ using the best fit 
constraint by \citet{kim02} $B=0.0032\;, \gamma=2.37$. The column density 
distribution is given by 
$f(N_{\mbox{\scriptsize\ion{H}{i}}}) \propto N_{\mbox{\scriptsize\ion{H}{i}}}^{-\beta}$ 
where the slope is $\beta\sim1.5$. The Doppler parameter distribution is given by
 $\mathrm{d}n / \mathrm{d}b \propto b^{-5}\mathrm{exp}\left[{-{b_{\sigma}^4}/{b^4}}\right]$ 
where  $b_{\sigma}\simeq 24\;\mathrm{km/s}$ \citep{kim01} is a parameter depending on 
the average amplitude of the fluctuations in the velocity space of the absorbers 
\citep{Hui99}. Each absorption feature was modelled as a Voigt profile and once a 
transmission spectrum was computed, we multiplied it by a QSO template spectrum as described in the next section.

%------------------------------------------------------------------------------
%Figure 
%----------------------------------------------------------------------------
\begin{figure}
\resizebox{\hsize}{!}{\includegraphics*[angle=-90]{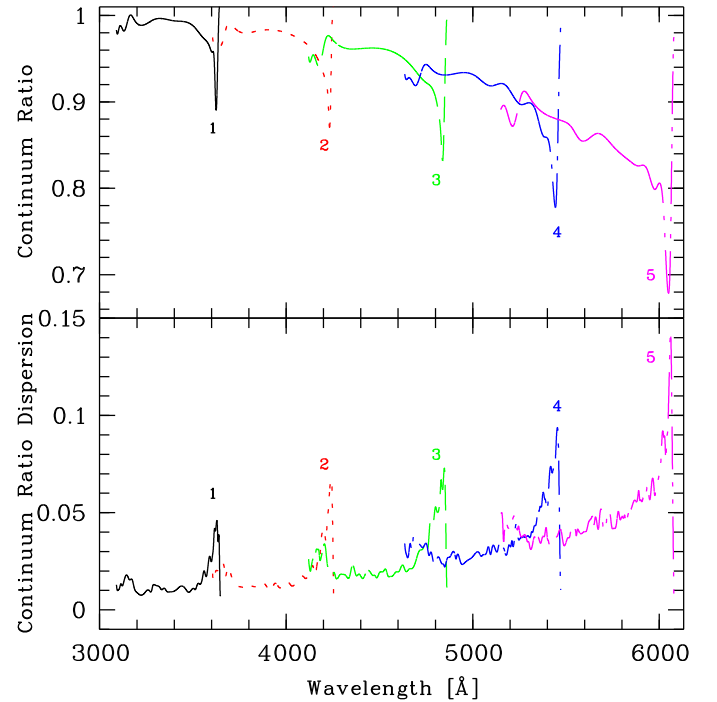}}
\caption{Top panel: Average ratio between the fitted and input continuum for the five sets of simulated QSOs. Bottom panel: Standard deviation profiles relative to the above systematic bias.}
\label{contchange}
\end{figure}
%------------------------------------------------------------------------------

%------------------------------------------------------------------------------
%Figure 
%----------------------------------------------------------------------------
\begin{figure}
\resizebox{\hsize}{!}{\includegraphics*[]{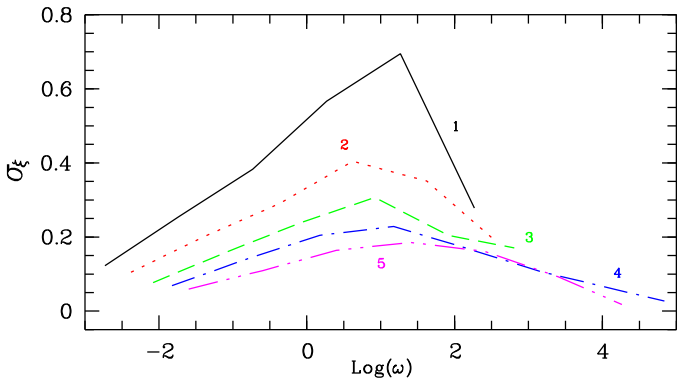}}
\caption{Statistical (shot noise) uncertainties of the normalised optical 
  depth close to a quasar, estimated from the Monte-Carlo simulations.}
\label{staterrlos}
\end{figure}
%------------------------------------------------------------------------------

%------------------------------------------------------------------------------
%Figure 
%----------------------------------------------------------------------------
\begin{figure*}
\sidecaption
\includegraphics*[width=12cm]{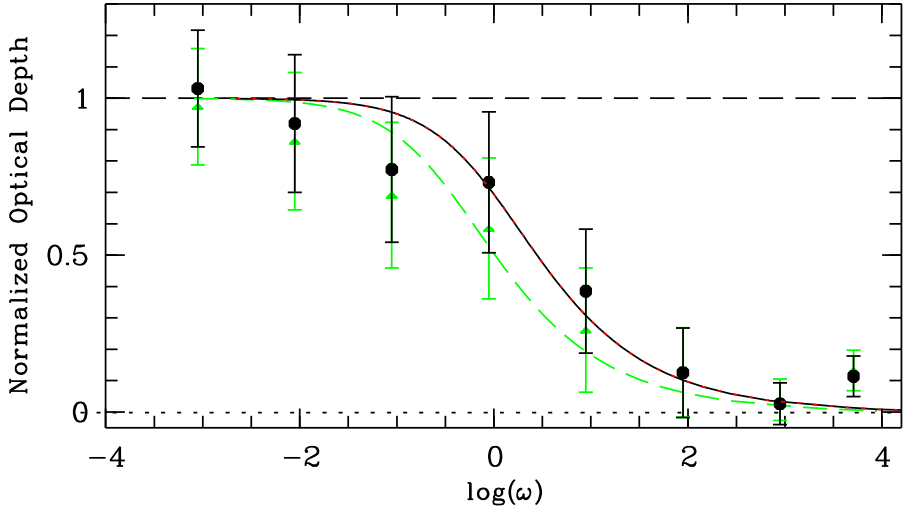}
\caption{Normalised Optical depth versus $\omega$ profile for the combined sample of 17
  quasars, binned in steps of $\Delta\log\omega = 1$. The signature of the
  proximity effect is clearly visible for both continuum corrected and uncorrected profiles. 
  The curved lines are the best fit of the simple photoionisation model to the data,
  corresponding to a UV background of $\log(J_\nu)=-21.03$ (solid line), in units of [erg cm$^{-2}$s$^{-1}$ Hz$^{-1}$ sr$^{-1}$], while the 
green long-dashed line correspond to the best fit to the continuum uncorrected data.
  The horizontal line refers to the case of no proximity effect.} 
\label{pe_res} 
\end{figure*}
%----------------------------------------------------------------

\subsubsection{Quasar SEDs}

We generated a set of 200 artificial quasar spectral energy
distributions (SEDs) via the principal component method as described by 
Suzuki (2006). Each rest-frame quasar spectrum can be decomposed as
$f_\lambda(\lambda) = \mu(\lambda) + \Sigma(c_\mathrm{i} \cdot p_\mathrm{i}(\lambda))$
with a mean spectrum $\mu$, the principal component spectra $p_\mathrm{i}$ and the
coefficients $c_\mathrm{i}$. We adopted the principal components by \citet{suzuki05}, 
who determined $\mu$ and $p_\mathrm{i}$ at $1020\ \mathrm{\AA}<\lambda_\mathrm{rf}<1600\ \mathrm{\AA}$ 
from 50 HST FOS spectra of low-redshift quasars. The coefficients $c_\mathrm{i}$ are
approximately Gaussian distributed \citep{suzuki06}. After generating the
200 SEDs by drawing the $c_\mathrm{i}$ from their Gaussian distributions, 
we convolved it with the instrumental profile and added random Gaussian 
noise to reproduce our observations.

In order to investigate how different QSO emission redshifts affected the error budget, 
we simulated quasars at five typical redshifts 
$z$ = 2.0, 2.5, 3.0, 3.5, 4.0, (denoted in Figs.\ 
\ref{contchange}--\ref{staterrlos} with numbers
1, 2, 3, 4 and 5, respectively) and luminosities 
representative for our sample.

We then normalised our spectra as performed for the actual observations. 
These simulations created a database
to quantitatively assess the two dominant sources of error: 
the limited number of absorbers per individual line of sight (cosmic
variance), and the misplacement of the continuum because of line
crowding. In the following we consider both error sources in turns.

\subsubsection{Systematic errors: Continuum placement}\label{syserrors}

For each simulated spectrum we computed the ratio between the fitted and the input
continuum. We then averaged the ratio over all
realisations at each emission redshift. The result is shown in the top panel of 
Fig.~\ref{contchange}, for the five redshifts adopted. Expectedly, the highest
deviation is always in the wings of the Ly$\alpha$ emission line.
It can also be seen that the increasing line number density with redshift 
causes a gradually growing systematic error. We use this profiles to correct our automatic 
continuum estimates and use the standard deviations shown in the bottom panel of 
Fig.~\ref{contchange} as contribution to the uncertainties. The effect of this correction 
can be seen in all the proximity effect plots as difference between the green triangles 
and the black dots in Fig.~\ref{pe_res}-\ref{PE_prof_sample}.

We note in passing that a second source of systematic errors would be the
presence of metal line systems in the proximity effect zone. Our spectral
resolution is insufficient to identify individual metal lines in the Lyman
forest, and any such absorption present, but unaccounted for, will
systematically increase the normalised optical depth $\xi$ and thus tend to
mask the proximity effect.

\subsubsection{Statistical errors}

We modelled the statistical error of the measured optical depth
along individual lines of sight as arising from Poissonian shot noise due
to the limited number of absorbers in each simulated spectrum. For each
stack of simulations we computed the mean and standard deviation of $\xi$
per $\log\omega$ bin. The results are shown in Fig.~\ref{staterrlos},
which demonstrates that the standard deviations $\sigma_\xi$ are always
considerably bigger than the continuum dispersion. However, recall that we 
fully account for both random and systematic errors. As expected,
the statistical errors become bigger towards lower redshifts due to
the smaller line number density. The simulation results were then used to
describe $\sigma_\xi(\omega,z)$ with a simple polynomial parameterisation.
Without introducing the proximity effect, the statistical error at high 
$\omega$ would be much larger due to cosmic variance on very small scale.

In order to estimate the statistical error for the combined analysis of
the full sample we ran a new set of simulations. Here we generated 10 random lines of
sight with the emission redshift of each of the 17 quasars in the sample,
and computed the statistical scatter after averaging over the 17
contributions to each $\omega$ value. We found $\sigma_\xi$
to be essentially independent of $\omega$ in this case.

\section{The proximity effect in the combined sample}

\subsection{A new estimate of the UV  background} \label{combined}

Figure~\ref{pe_res} summarises the main results regarding the 
combined sample of 17 QSO spectra. We see a highly significant reduction of
normalised optical depths at $\log\omega \ga 0$, i.e.\ for the zone
where photoionisation due to the local quasar-induced radiation field
is expected to prevail over the metagalactic UV background. As already
demonstrated by BDO88, this turnover can be used to constrain
the mean intensity of the UVB. We adopted the fitting formula
\begin{equation}\label{eq_res}
F(\omega)=\left(1+\frac{\omega_{-21}}{a}\right)^{1-\beta}
\end{equation}
with $a$ being the free parameter and $\omega_{-21}$ is the value
of $\omega$ relative to a reference value of the UVB,  
$J^\star_{-21} \equiv 10^{-21}$ erg cm$^{-2}$s$^{-1}$ Hz$^{-1}$ sr$^{-1}$.
We then applied a straightforward $\chi^2$ minimisation to search 
for the best-fitting value of $a = J(\nu_\mathrm{LL})/J^\star_{-21}$.

In doing this computation we found that the bin size $\Delta\log\omega$ has
some moderate effect on the resulting best-fit value of the UVB.  If the data 
are merged with very small or even without any binning, $a$ will be biased
towards low values because of the substantial scatter in $\xi$ at very small
$\omega$ (for $\log(\omega)\la -1.5$), due to the strong effects of shot
noise for this narrow $\log(\omega)$ interval. 
On the other hand, too large bins will tend to hide the signature of the 
proximity effect, thus make the UVB appear stronger than it really is.
As a reasonable compromise we chose $\Delta\log\omega = 1$, upon which
also Fig.~\ref{pe_res} is based. 
An additional effect which tend to change the estimation of the UVB is the 
normalisation used to compute $\xi$ (see eq.~\ref{eq:xi}). We address this 
problem with two different strategies. We use the normalisation by \citet{kim02} 
and our normalisation ($B=0.0041$) to reach $\xi_{\omega \rightarrow 0} \sim 1$ 
for the combined set of Ly$\alpha$ forest regions. Tab.~\ref{Tab3} summarises our 
results revealing a maximal dispersion of about $\sim 0.9$. We decide to use $B=0.0041$ 
for our results. 

As best fit value we obtain 
$J(\nu_\mathrm{LL}) = 9 \pm 4 \times 10^{-22}$~erg cm$^{-2}$ s$^{-1}$
Hz$^{-1}$ sr$^{-1}$, or in logarithmic
units, $\log J(\nu_\mathrm{LL}) = -21.03^{+0.15}_{-0.22}$.
Using a slightly narrower bin size of $\Delta\log\omega = 0.7$
lowers $J(\nu_\mathrm{LL})$ by about 0.05~dex.

This estimate of the UVB intensity is in very good agreement with all 
recent measurements based on a wide range of techniques and data sets. 
For example, \citet{scott00} obtained 
$J(\nu_\mathrm{LL})/J^\star_{-21} = 0.7_{-0.44}^{+0.34}$, applying
line count statistics on more than hundred spectra at $\sim 1\AA$ resolution.
More similar to our approach, \citet{liske01} used the flux transmission
statistic on 10 QSO spectra with $\sim 2$~\AA\ resolution and a S/N of $\sim
40$, obtaining $J(\nu_\mathrm{LL})/J^\star_{-21} = 0.35_{-0.13}^{+0.35}$.
Not much has yet been published using high resolution spectra.
\citet{giallongo96} obtained $J(\nu_\mathrm{LL})/J^\star_{-21} =
0.5\pm 0.1$ and \citet{cooke97} $J(\nu_\mathrm{LL})/J^\star_{-21} =
0.8_{-0.4}^{+0.8}$, again close to our value even though they are all smaller  
(up to a factor of about 2).
This order of magnitude is also consistent with predictions based on the QSO
luminosity function \citep{haardt96}.

\subsection{Dependence on model parameters}

Concerning the evolution of $\tau_\mathrm{eff}$ in the Ly forest (see eq.~\ref{eq:tau}), 
we regard only the normalisation $B$ as variable. The slope $\gamma$ has been estimated by 
several authors with much higher accuracy in the past years \citep[e.g.,][]{kim01,kim02,schaye03}. 
We consider only two normalisations: a conservative one with $B=0.0032$ \citep{kim02}, which 
returns a higher UVB estimate since $\xi_{\omega \rightarrow 0} \gtrsim 1$, and $B=0.0041$ 
which we adopt since it leads to $\xi_{\omega \rightarrow 0} \simeq 1$. We believe that the 
discrepancy between our adopted value and the value by \citet{kim02} is due to a combination 
of line blending and resolution effects. Our results are sensitive to the slope of the column 
density distribution $\beta$ which sets the steepness of $\xi(\omega)$ in the BDO88 ionisation 
model. The column density distribution is very well approximated by a single power law with 
$\beta \simeq 1.5$ over 10 dex \citep{petitjean93}. However, it has been shown that the slope 
changes somewhat with redshift \citep{kim02}. Since the lines are unresolved in our low-resolution 
spectra, we fix a single power law distribution and vary its assumed value by $\Delta \beta = 0.1$ 
to estimate the robustness of our UVB measurements. The Doppler parameter distribution has no 
direct impact on our results since the BDO88 model simply assumes an isothermal \ion{H}{i} distribution.\\
Table~\ref{Tab3} summarises the dependence of our UVB estimates on model parameters. There is a 
substantial scatter in the estimations and both $B$ and $\beta$ play a central role. The binsize 
has only a small (but still detectable) effect.

%---------------------------------------------------------------------------
%Table 
%---------------------------------------------------------------------------
\begin{table}
\small\centering
\caption{Estimations of the UV background $J(\nu_\mathrm{LL})/J^\star_{-21}$ (computed in units of erg cm$^{-2}$s$^{-1}$ Hz$^{-1}$) from different model parameters ($\beta$ and $B$) and binsizes $\Delta\log\omega$.}
\label{Tab3}
\begin{tabular}{lllll}
\hline\hline
\noalign{\smallskip}
 & \multicolumn{2}{c}{$B=0.0032$} & \multicolumn{2}{c}{$B=0.0041$}\\
$\Delta\log(\omega)$ & $0.7$&$1.0$ & $0.7$& $1.0$\\
\noalign{\smallskip}\hline\noalign{\smallskip}
$\beta = 1.4$ &$1.01 $ &$1.12 $ &$0.40 $ &$0.51 $ \\
$\beta = 1.5$ &$1.89 $ &$2.04 $ & $0.88 $& $0.93 $\\
$\beta = 1.6$ &$2.83 $ &$3.01 $ & $1.19 $& $1.41 $\\
\noalign{\smallskip}
\hline
\end{tabular}
\end{table}
%---------------------------------------------------------------------------

%------------------------------------------------------------------------------
%Figure 
%----------------------------------------------------------------------------
\begin{figure*}
\resizebox{\hsize}{!}{\includegraphics*{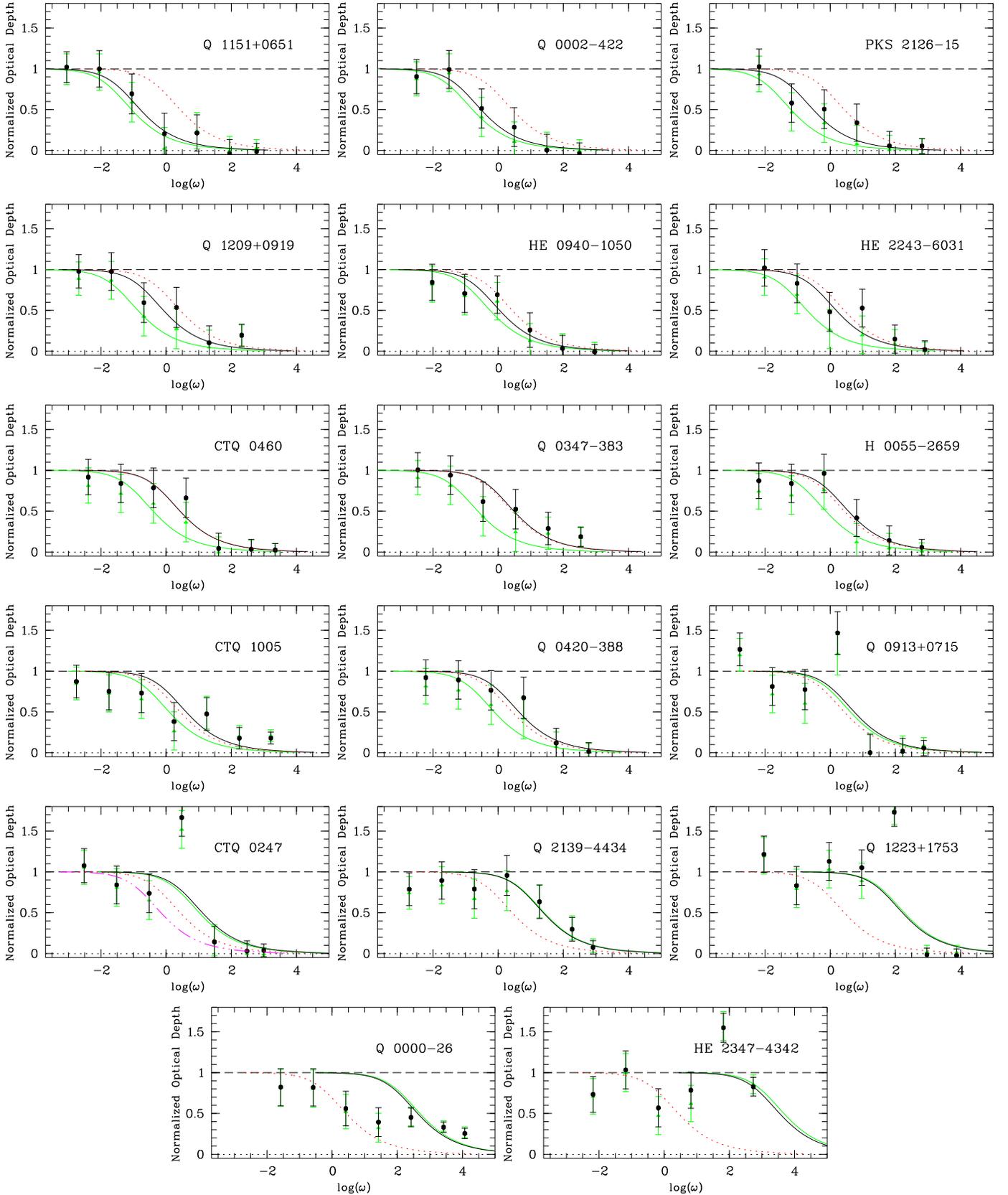}}
\caption{Search for proximity effect signatures in individual lines of sight,
  for all 17 QSO spectra. Each panel shows the normalised optical depth $\xi$
  versus $\omega$ in the same way as Fig.~\ref{pe_res}, with the best-fit
  model of the combined analysis superimposed as dotted red lines. The solid lines
  delineate the best fit to each individual QSO as described in the text.
  For CTQ 0247 we also plot the best fit
  excluding the strong absorption at $z\sim 3.014$ (purple curve). The panels
  are sorted in order of decreasing strength of the proximity effect
  (horizontal displacement of solid and dotted lines). HE~2347$-$4342 has no
  detectable proximity effect in \ion{H}{i}.}
\label{PE_prof_sample}
\end{figure*}
%------------------------------------------------------------------------------

\section{The proximity effect in individual lines of sight}\label{indiv_qsos}

\subsection{Results}

The proximity effect is generally seen as a statistical phenomenon, which may
or may not be detectable in individual spectra. The high S/N of our spectra
motivated us to search for proximity effect signatures in each of our 17
quasar spectra. The basic approach was essentially identical to that of the
combined analysis. Compute the mean normalised optical depth for a given line
of sight within a given bin $\Delta\log\omega$ and check whether $\xi$
systematically decreases for large values of $\omega$. 
We set the same normalisation for every object as for the combined analysis.
The $\omega$ scale was
now fixed by assuming the value of the mean UV background intensity from the
combined analysis.  The results are displayed in Fig.~\ref{PE_prof_sample},
one small panel per quasar.  The error bars are now of course dominated by
Poissonian shot noise, estimated from the simulations as described above.
In each panel the expected run of $\xi$ with $\omega$ is shown as the 
dotted red line, assuming that the metagalactic UV background is constant 
and has the same spectral shape as the quasar. We also show the profile before 
the systematic continuum correction and its best fit model (green triangles and line).

Figure~\ref{PE_prof_sample} demonstrates that in all except one case, $\xi$
decreases substantially from left to right. Thus we can say that the proximity
effect is detected in 16 out of our 17 quasar spectra. In the majority of
spectra, the $\xi$-$\omega$ profile is even formally consistent with the
prediction based on the combined analysis. In a number of cases the data 
seem to be (mildly) discrepant with the prediction; we discuss some of 
these cases below.

We also applied the above fitting procedure to each spectrum separately; the
results of that exercise are shown as the solid curves in
Fig.~\ref{PE_prof_sample}.  Table~\ref{Tab4} summarises the fit results.
Apparently the value of the fitting parameter $a$ shows
significant scatter between the different quasars. The value of $a$ describes the 
horizontal offset of the solid curve relative to the dotted curve and in the following is regarded 
 as a quantitative measure of the strength of the 
proximity effect signal (in the sense that a large $a$ means a weak 
proximity effect).
This does of course not 
imply that the UV background fluctuates by a similar amount. 
While statistical errors certainly contribute to the scatter, one may also
reinterpret the parameter $a$ as a measure of the flux of the quasar
at the Lyman limit. This flux may even not always have been constant over
the light travel time across the proximity effect zone; we return to that
point in Sect.~\ref{lifetime} below. 

We now briefly comment on two objects where the proximity effect appears to
be extremely weak or absent. In the case of CTQ 0247, there is a strong
associated absorption system at $z_\mathrm{abs} = 3.014$, corresponding to 
$\log \omega \lesssim 1$. Removing this absorption manually from the spectrum
and redoing the analysis yielded a proximity effect signal perfectly
consistent with the prediction from the combined analysis.

\subsection{A hidden proximity effect for HE 2347$-$4342?}\label{strongabs}

For HE~2347$-$4342, we see no evidence at all of a downturn of
$\xi(\omega)$. The absence of any proximity effect was noticed already by
\citet{reimers97} upon mere visual inspection of the spectrum. Again there is
a conspicuous strong associated absorption system. Although in this case a
removal of that system does not dramatically improve the proximity effect
signal, the absorber may nevertheless play an important role in this line of
sight, as it may attenuate the ionising radiation field towards
lower redshifts. This is supported by the following simple calculation.

The total measured \ion{H}{i} column density in the associated system is of
the order of $2\times10^{16}$~cm$^{-2}$ \citep{fechner04}. We assume this to be
located in an absorbing slab of matter very close to the QSO. Knowing the
Lyman limit flux of the QSO, we can predict the value of $\omega$ at the
location of the slab, $\log\omega\simeq 2.1$; this immediately relates to a
predicted reduction of \ion{H}{i} column density in the slab of $\sim
2$~dex. Thus, the column density of the same absorber without the QSO ionising
radiation would be of the order of $10^{18}$~cm$^{-2}$. This would be
sufficient to render a remaining proximity effect for the line of sight
undetectable.

We note that \citet{fechner04} detected evidence of a hard radiation field
from a detailed photoionisation modelling analysis of metal absorption lines
in this system. A similar trend is apparent in our recent investigation of
the \ion{He}{ii} Lyman forest, combining VLT and FUSE high-resolution spectra 
of this quasar \citet{worseck07}.  Thus, while the traditional \ion{H}{i}
proximity effect is clearly absent in HE~2347$-$4342, most probably due
to excess absorption close to the QSO, there are clear signs of a
`proximity effect in spectral hardness' (cf.\ \citealt{worseck06}) for this
object. We therefore conclude that all 17 quasars in our sample show
evidence of a genuine proximity effect.

%---------------------------------------------------------------------------
% Table
%---------------------------------------------------------------------------
\begin{table}
\centering
\caption{Input and results used for the photoionisation model fits.
  $f_{\mathrm{LL}}$ is the predicted Lyman limit flux of each quasar.
  In the treatment of individual lines of sight, $\log(a)$ is the 
  fitted parameter used to quantify the strength of the proximity
  effect. CTQ~0247 is listed twice, with and without the associated
  absorption system.}
\label{Tab4}
\begin{tabular}{lcccccc}
\hline\hline\noalign{\smallskip}
QSO& $z$ & $\log(f_{\mathrm{LL}}(0))$ & $\log(a)$ \\
\noalign{\smallskip}\hline\noalign{\smallskip}
% BIN 1.0
Q 1151$+$0651  & 2.758 & $-27.00^{+0.01}_{-0.01}$ & $-1.11^{+0.51}_{-\infty}$ \\
Q 0002$-$422   & 2.767 & $-26.46^{+0.05}_{-0.06}$ & $-0.84^{+0.49}_{-\infty}$ \\
PKS 2126$-$15  & 3.285 & $-26.42^{+0.04}_{-0.05}$ & $-0.84^{+0.41}_{-\infty}$ \\
Q 1209$+$0919  & 3.291 & $-26.92^{+0.01}_{-0.01}$ & $-0.40^{+0.44}_{-\infty}$ \\
HE 0940$-$1050 & 3.086 & $-26.16^{+0.03}_{-0.03}$ & $-0.27^{+0.48}_{-\infty}$ \\
HE 2243$-$6031 & 3.010 & $-26.11^{+0.02}_{-0.02}$ & $-0.14^{+0.46}_{-\infty}$ \\
CTQ 0460       & 3.139 & $-26.55^{+0.03}_{-0.02}$ & $+0.11^{+0.41}_{-\infty}$ \\
Q 0347$-$383   & 3.220 & $-26.67^{+0.01}_{-0.01}$ & $+0.16^{+0.46}_{-\infty}$ \\
H 0055$-$2659  & 3.665 & $-26.61^{+0.03}_{-0.03}$ & $+0.24^{+0.43}_{-\infty}$ \\
CTQ 1005       & 3.205 & $-26.95^{+0.03}_{-0.03}$ & $+0.29^{+0.47}_{-\infty}$ \\
Q 0420$-$388   & 3.120 & $-26.37^{+0.03}_{-0.03}$ & $+0.31^{+0.44}_{-\infty}$ \\
Q 0913$+$0715  & 2.787 & $-26.74^{+0.02}_{-0.02}$ & $+0.42^{+0.50}_{-\infty}$ \\
CTQ 0247       & 3.025 & $-26.62^{+0.02}_{-0.03}$ & $+0.70^{+0.40}_{-\infty}$ \\
CTQ 0247       &       &                          & $-0.46^{+0.49}_{-\infty}$ \\
Q 2139$-$4434  & 3.214 & $-26.93^{+0.01}_{-0.02}$ & $+1.06^{+0.42}_{-\infty}$ \\
Q 1223$+$1753  & 2.945 & $-27.10^{+0.01}_{-0.01}$ & $+2.29^{+0.44}_{-\infty}$ \\
Q 0000$-$26    & 4.098 & $-26.19^{+0.04}_{-0.05}$ & $+2.30^{+0.24}_{-\infty}$ \\
HE 2347$-$4342 & 2.885 & $-26.20^{+0.03}_{-0.03}$ & $\gtrsim +3 $             \\
\noalign{\smallskip}\hline
\end{tabular}
\end{table}
%---------------------------------------------------------------------------

\subsection{Variations in the strength of the proximity effect}\label{2powlaw}

We now investigate how much of the scatter in the fitting parameter $a$ 
 might be attributed to uncertainties, or shot noise, or intrinsic dispersion
of other relevant properties.

\begin{figure}
\resizebox{\hsize}{!}{\includegraphics*[ angle=-90]{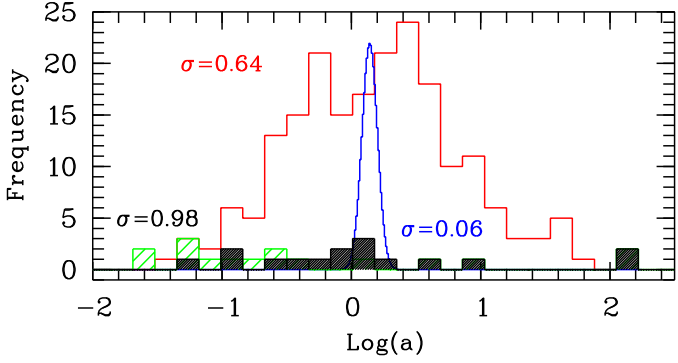}}
\caption{Distribution of the parameter $\log(a)$ for the fits to the 17
  individual spectra with and without continuum correction (black and green hashed histogram respectively). 
Overplotted is the distribution
  resulting only from shot noise in the simulated spectra (solid line),
  and the expected distribution of $\log(a)$ due to dispersion of spectral 
  indices (blue gaussian).}
\label{sigma_comp}
\end{figure}

Figure~\ref{sigma_comp} shows that formally the best-fit values of $a$, before and after 
continuum correction (green and black histograms respectively),
extend over several orders of magnitudes, with a standard deviation of
$\sigma_{\log(a)}\sim 1$ (black histogram).

The large dispersion can to a small part be explained by uncertainties in quasar magnitudes and redshifts, which both affect the computation of $\omega$. For those objects where we have rather accurate V magnitudes (with errors $\le 0.1$ mag), the flux scale is accurate to within 9\%. For the remaining objects the flux uncertainties might be as large as 30\%. The direct effect on $\log(a)$ is an uncertainties of 0.03--0.1 dex. In addition, redshift errors (Tab.~\ref{Tab1}) might shift $\log(a)$ by 0.05--0.1 dex. We conclude that these uncertainties cannot be the main source of spread in the estimated $\log(a)$ values.

We employed our Monte-Carlo simulations to 
estimate the expected scatter solely due to statistical shot noise errors
(i.e., cosmic variance). To this effect we first systematically reduced
the optical depths in the simulated spectra following strictly the theoretical
proximity effect prescription (Eq.~\ref{eq:xi}), after which we 
`remeasured' the proximity effect and its strength $a$ by fitting
the artificial data in the same way as the observed ones. The resulting
histogram of $a$ values is superimposed in Fig.~\ref{sigma_comp}; the
distribution is approximately Gaussian with a standard deviation of 
$\sigma_{\log(a)}=0.64$. This is substantial and implies that 14 our of
17 of our quasars are located within $\pm 2\sigma$ expected for pure random 
errors. Nevertheless, 3 quasars from
our sample are located outside of this interval, which is a bit much to
declare them all as outliers.

This means that other effects play a role, and that these effects may lead to 
gross deviations from the simple expectation. Possible mechanisms
might be, for example, a dispersion in spectral indices for the quasars, 
long time scale variability of the quasars, or a strongly fluctuating UV
background. We first consider the effects of non-uniform spectral indices.

In the standard analysis of the proximity effect as introduced by BDO88,
the assumption of a uniform quasar spectral index is obviously wrong, but
the averaging over samples of quasars makes the analysis very insensitive
to any intrinsic dispersion. When considering individual quasar lines of
sight as we do here, this assumption may be more harmful. We investigated
this by extending the original calculation by BDO88,
solving the exact integrals for the ratio of the
photoionisation rates of the quasar and the background and using the
photoionisation cross section for a pure hydrogen cloud, so that 
\begin{equation}
\omega = \frac{\int_{\nu_{\mathrm{LL}}}^{\infty}{\frac{4\pi J^{\mathrm{q}}_{\nu}\left(\nu\right)}{h\nu}\,\sigma\left(\nu\right)\mathrm{d}\nu}}{\int_{\nu_{\mathrm{LL}}}^{\infty}{\frac{4\pi J^{\mathrm{b}}_{\nu}\left(\nu\right)}{h\nu}\,\sigma\left(\nu\right)\mathrm{d}\nu}} = \omega_{\mathrm{old}}\frac{\kappa-3}{\alpha-3}\label{pli1}
\end{equation}
where $\omega_{\mathrm{old}}$ is the know expression of eq.~\ref{34.6} and
$\kappa \sim -1.8$ is the background spectral index following the model by
\citet{haardt96}. Expressing everything in logarithmic units we arrive at 
\begin{equation}\label{pli2}
\log(\omega) = \log(\omega_{\mathrm{old}}) + \sigma_{\log(a)}\:.
\end{equation}
Real quasars show a dispersion of spectral slopes of $\sigma_\alpha = 0.25$
around $\overline{\alpha}= -0.46$ (derived from the SDSS QSO composite spectrum
and its standard deviation published by \citealt{berk01}). The values of $\omega$ are
therefore offset (i) by $\Delta\log(\omega)=0.142$ on average, and (ii) by a
random component with a standard deviation of $\sigma_{\log(a)}=0.06$
(included as the narrow Gaussian in Fig.~\ref{sigma_comp}). Obviously,
the dispersion in spectral slopes is too small, by a long way, to account
for the observed distribution of $\log(a)$.

Among the two other mentioned options, intrinsic fluctuations of the UV
background might also contribute, but numerical simulations suggest that 
substantial fluctuations are only expected for redshifts higher than those
covered here \citep{Croft99, McDonald05}. 

The presence of metal transitions falling in the proximity effect zone or a still imperfect
 continuum placement might also enhance the scatter.

A very plausible effect,
on the other hand, would be significant QSO variability over timescales 
of the light travel time across the proximity effect zones which is of the
order of $10^7$ years; there is no reason to expect that quasars always 
maintain their radiative output over such long periods. If they do not,
then the observed Lyman limit luminosity will not be the same as that 
received by clouds along various points along the line of sight. We will
discuss a specific aspect of this effect in the next section.

%------------------------------------------------------------------------------
%Figure 
%----------------------------------------------------------------------------
%------------------------------------------------------------------------------

%------------------------------------------------------------------------------
%Figure 
%----------------------------------------------------------------------------
\begin{figure}
\resizebox{\hsize}{!}{\includegraphics*[]{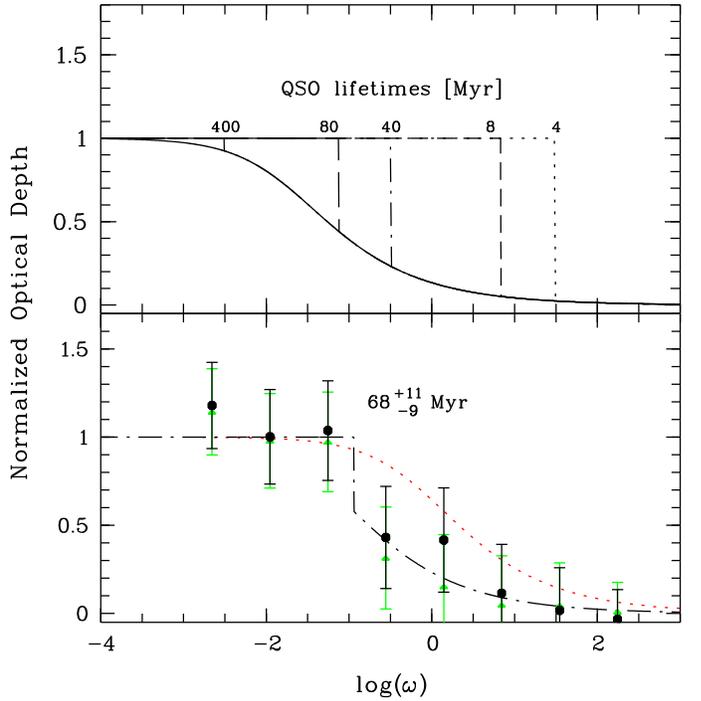}}
\caption{Top panel: Expected $\xi(\omega)$ model for different finite
  lifetimes of the quasar. The solid curve represents the BDO88 model 
  of the proximity effect, with a QSO of infinite lifetime.
  Bottom panel: Tentative application of the above model to the QSO
  Q~0002$-$422. The dotted line is the predicted $\xi(\omega)$ curve
  for infinite lifetime, the dot-dashed line delineates the curve
  for a fiducial finite lifetime of 68 Myr.}
\label{lifemodel}
\end{figure}
%----------------------------------------------------------------------------

\subsection{Finite quasar age} \label{lifetime}

The simple theory of the proximity effect implicitly includes the assumption
that quasars shine for an infinite time, or at any rate for much longer than
the light crossing time of the proximity effect zone. This may be wrong,
and we now ask specifically what would happen if a quasar is abruptly
switched on within, say, less than a few Myrs before the observation.
As soon as the quasar starts to radiate in the UV, an over-ionised 
``sphere'' will start to expand around the quasar (always assuming spherical
symmetry). The new equilibrium state with ionising photons from the local
source and the UV background -- requires at least some $\sim 10^4$~yrs
to establish, which means that for even younger QSOs the proximity effect will
be absent. 

The light travel time between two redshifts along a single line of sight can
be calculated for a $\Lambda$-Universe as
\begin{equation}
t(z)=\frac{2}{3H_0\sqrt{\Omega_{\mathrm{\Lambda}}}}\mathrm{asinh}{\sqrt{\frac{\Omega_{\mathrm{\Lambda}}}{\Omega_{\mathrm{m}}(1+z)^3}}}
\end{equation}
\citep{peacock}, allowing us to convert redshift intervals into light travel
time differences.
Switching a quasar on can be expressed in the model by introducing a step
function into the $\omega$ profile,
\begin{equation}
\omega(z)=\left\{\begin{array}{ll}
0 & \mathrm{for}\:z<z_{\mathrm{life}}\\
\frac{f_{\nu}(\lambda_\mathrm{LL}(1+z_{\mathrm{c}}))}{4\pi J_{\nu}}\frac{1}{(1+z_{\mathrm{c}})}\left(\frac{d_{L}(z_{\mathrm{q}},0)}{d_{L}(z_{\mathrm{q}},z_{\mathrm{c}})}\right)^{2} &
 \mathrm{for}\:z_{\mathrm{life}}<z<z_{\mathrm{em}}\:.
\end{array}\right.
\end{equation}
Far away from the QSO (low $\omega$ values), the QSO has no effect yet.
When passing into the over-ionised region, the theoretical profile
$\xi(\omega)$ drops abruptly, joining the BDO88 prescription of the
proximity effect. This is shown in the top panel of Fig.~\ref{lifemodel},
where we present the expected profiles for different assumed ages. 
For presentation purposes, this model was computed for a QSO at $z_\mathrm{q}=2.76$ and
$f_\nu(\lambda_\mathrm{LL}(1+z_{\mathrm{q}}))=3.46\cdot10^{-27}$ erg cm$^{-2}$s$^{-1}$ Hz$^{-1}$.

It turns out that observable signatures can be expected for an interesting 
range of ages. Up to $t_\mathrm{QSO} \sim 4$ Myr the turnover happens so close to 
the quasar that even for a very luminous QSO it will be hard to detect, 
given the inevitable shot noise limitations. 
Above of at most $\sim 500$ Myr, on the other hand, the differences between
the models with and without lifetime will wash out completely because
$\xi(\omega)$ is expected to be close to unity anyway. The range between these
extremes, $4\:\mathrm{Myrs} \la t_\mathrm{QSO} \la  400\:\mathrm{Myrs}$,
is interestingly close to quasar lifetimes estimated by other, usually
much more indirect methods, such as modelling of QSO accretion 
\citep{yu02,hopkins06} or analysis of QSO clustering \citep{croom05}.
It might thus be possible to detect age effects of quasars by studying their
proximity effect signature.

We searched our quasars for possible examples of such a step feature, and
found one (highly tentative) possible example: Q~0002$-$422, shown in the
bottom panel of Fig.~\ref{lifemodel}, seems to match the expected pattern for
a `recently born quasar' quite well. The normalised optical depth $\xi$
remains constant at $\sim$ unity over the Lyman forest until
$\log(\omega)=-1.5$ is reached, where it rather abruptly drops to much lower
values.  We chose a smaller binning in $\log\omega$ for this plot, compared to
the previous figures, in order to highlight the relatively sudden drop.
Applying our simple model we obtain an age of $t_\mathrm{QSO} \simeq
68^{+11}_{-9}$ Myrs, where the errors are based on simply assuming the
uncertainty to be $\pm 0.35$ bin in $\log\omega$.  Of course we do not claim to
have measured the age of that quasar with 20~\% accuracy. The difference in 
the goodness-of-fit between finite and infinite quasar ages is not formally significant.  
But the exercise
shows that it is possible to derive empirical constraints on quasar ages from
proximity effect signatures, which might even be enhanced by using data of
higher spectral resolution.

\section{Conclusions}\label{conclus}

We have presented new evidence of the line-of-sight proximity effect as a
universal phenomenon occurring in the spectra of high-redshift quasars.  Even
though our spectra are limited in spectral resolution, their high S/N and the
power of the flux transmission method has enabled us to demonstrate the
presence of the effect for every single line of sight, for the first time.

Our estimate of the mean UV background intensity for the redshift range 
$2.7<z<4$ is $\log(J_\nu)=-21.03^{+0.15}_{-0.22}$, in very good agreement 
with literature values for similar redshift ranges. We made a careful assessment
of the error budget using extensive Monte-Carlo simulations. The errors are
clearly dominated by cosmic variance, which implies that better spectral 
resolution would not necessarily have a dramatic impact on the measurement
accuracy. Of course, high resolution spectra would be advantageous for 
a more detailed analysis of several other aspects of the proximity effect,
such as the effects of gravitational clustering of absorbers near the
QSO \citep{rollinde05}, which may overestimate the value of the UV 
background by up to a factor of 3 \citep{loeb95}.

We have quantified the strength of the proximity effect in individual
spectra and find that this shows a higher dispersion than expected from
only statistical shot noise errors. Among the most likely contributors
for this additional dispersion are again gravitational clustering of
absorbers near the QSO, or QSO variability over very long timescales;
fluctuations of the UV background are also possible but unlikely to
play a major role at this redshift range.
We presented a speculative, but conceptually simple observational
test to search for signatures of finite quasar ages in the optical
depth profiles derived for a single quasar, and we even tentatively
identified a candidate where such a pattern might be visible.

%------------------------------------------------------------------------------
\begin{acknowledgements}
 We would like to thank Prof. Piero Rafanelli and Dr. Stefano Ciroi for all the 
inspiring discussions and support. ADA is grateful to the Fondazione Ing. Aldo Gini 
and the University of Padova for providing grants for this work. ADA and GW acknowledge 
support by a HWP grant from the state of Brandenburg, Germany.
\end{acknowledgements}
%------------------------------------------------------------------------------

\bibliography{bibliography}
\Online
\begin{appendix}
\begin{figure}
\resizebox{\hsize}{!}{\includegraphics*[]{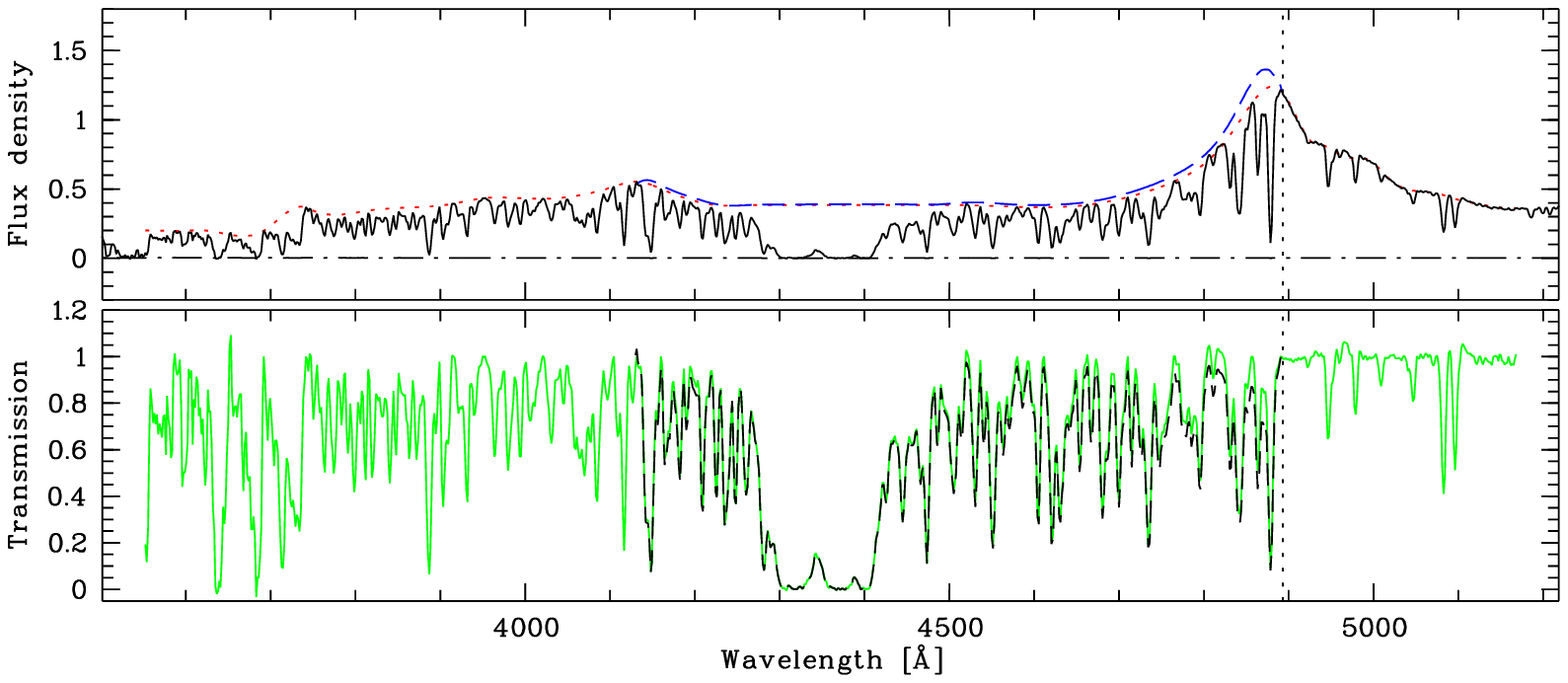}}
\caption{CTQ 0247 }
\end{figure} 
\begin{figure}
\resizebox{\hsize}{!}{\includegraphics*[]{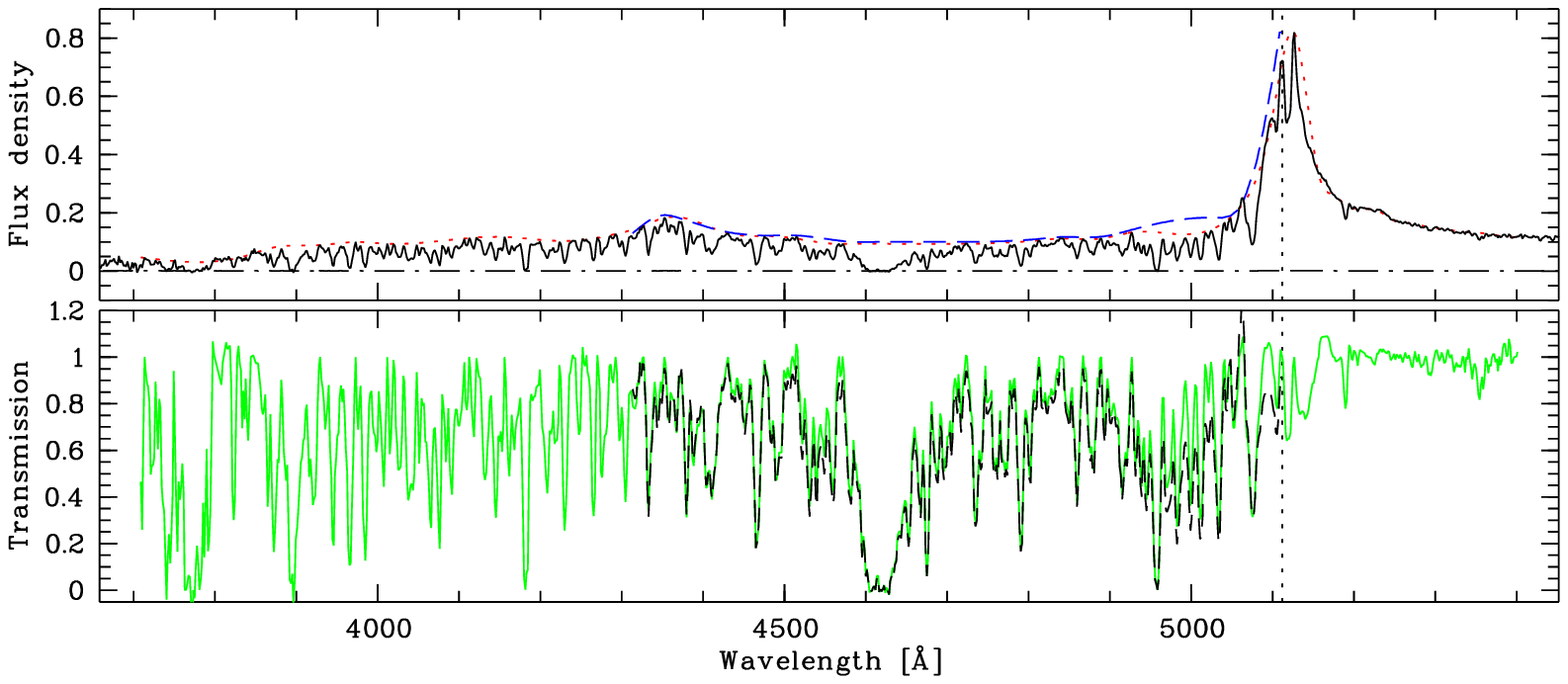}}
\caption{CTQ 1005}
\end{figure} 
\begin{figure}
\resizebox{\hsize}{!}{\includegraphics*[]{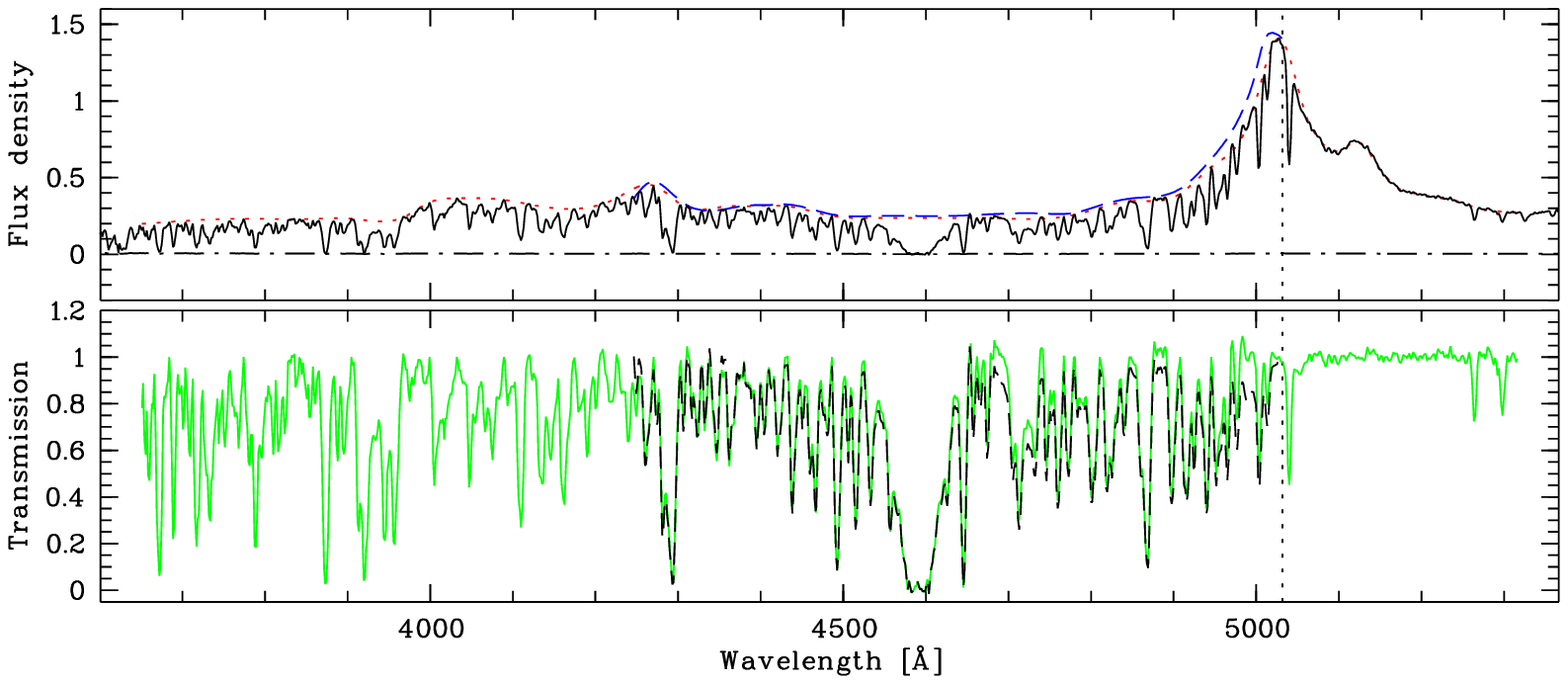}}
\caption{CTQ 0460}
\end{figure}     
\begin{figure}
\resizebox{\hsize}{!}{\includegraphics*[]{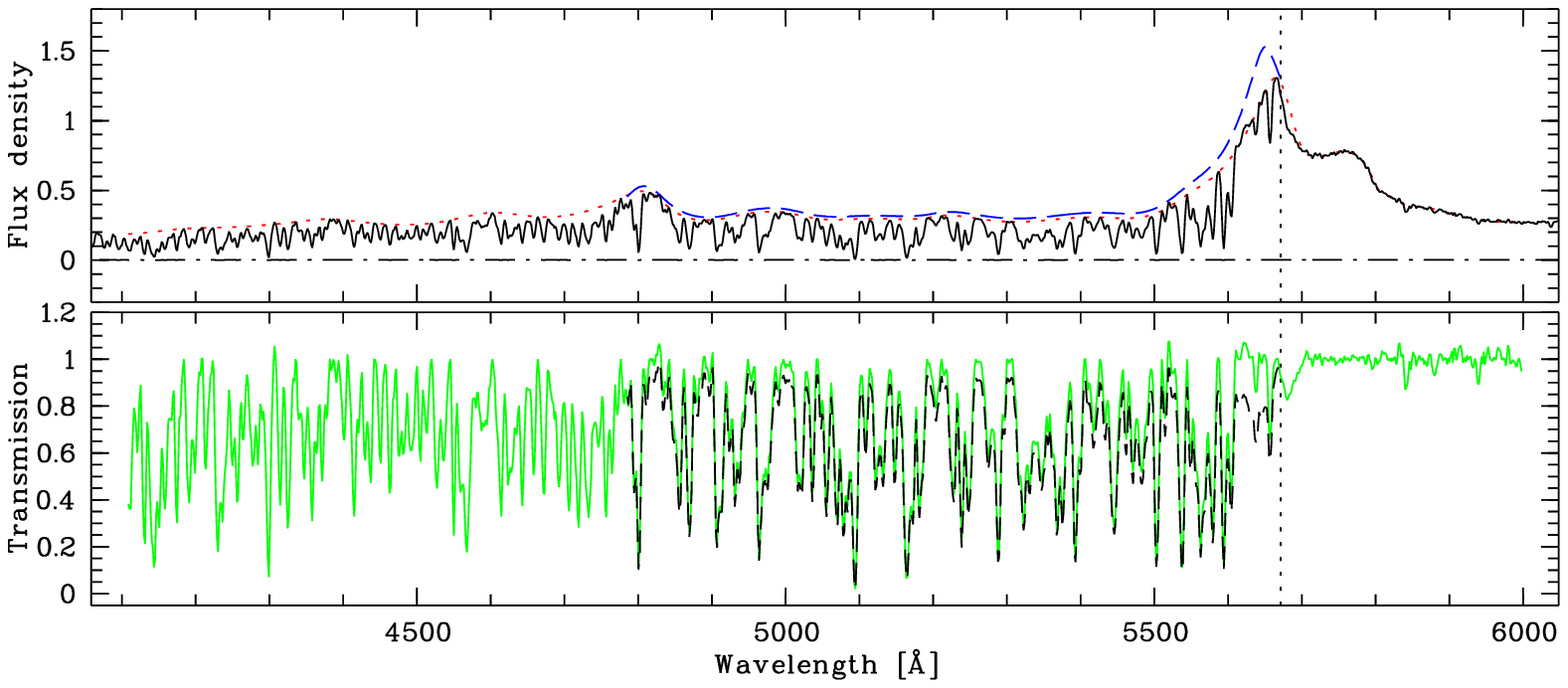}}
\caption{H 0055$-$2659}
\end{figure}      
\begin{figure}
\resizebox{\hsize}{!}{\includegraphics*[]{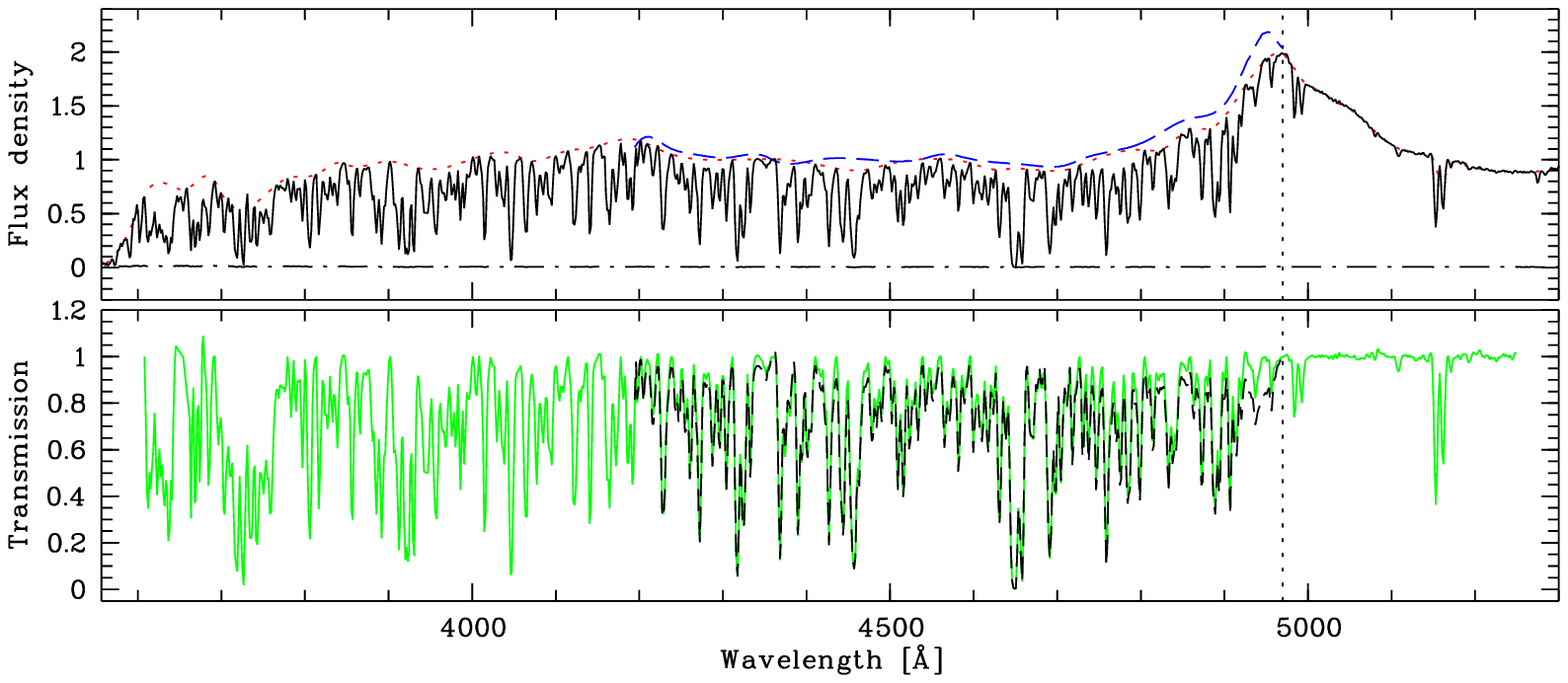}}
\caption{HE 0940$-$1050}
\end{figure}     
\begin{figure}
\resizebox{\hsize}{!}{\includegraphics*[]{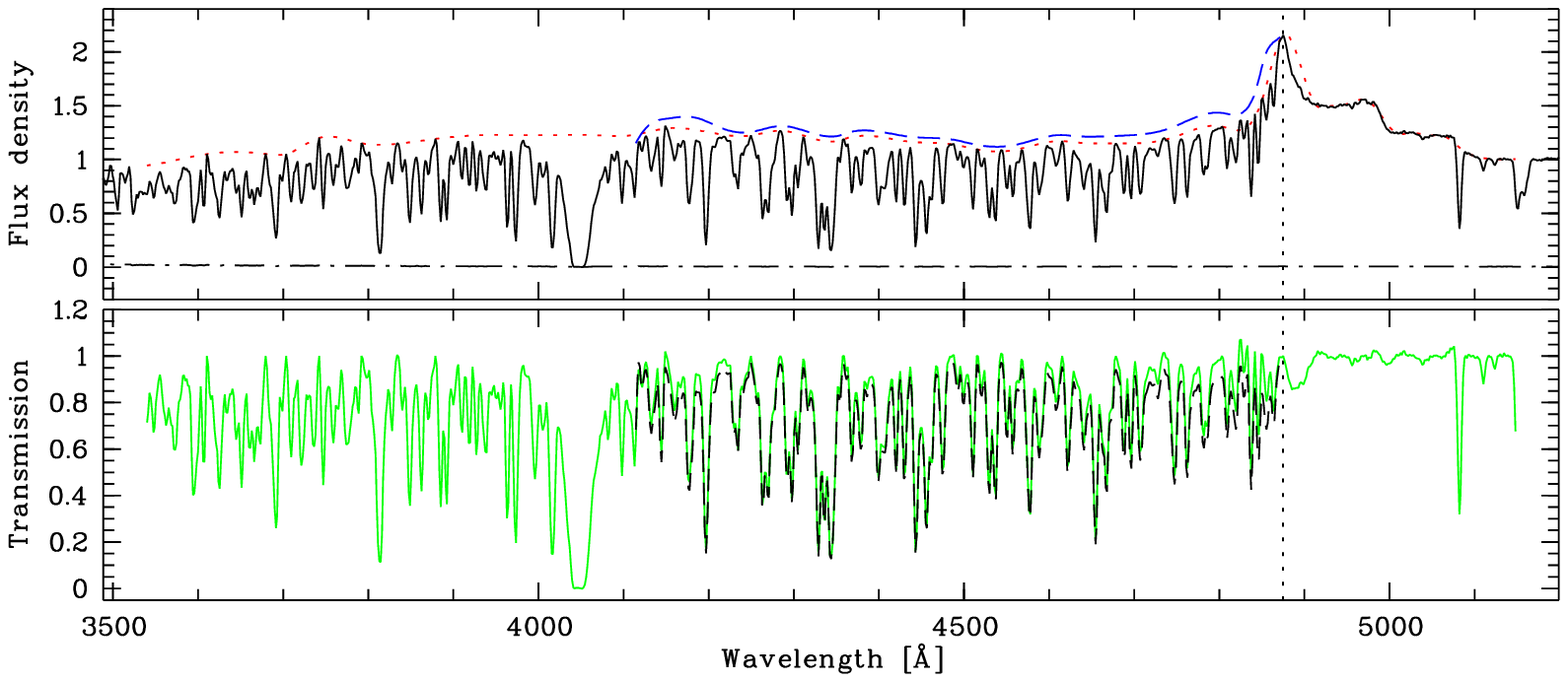}}
\caption{HE 2243$-$6031}
\end{figure}  
\begin{figure}
\resizebox{\hsize}{!}{\includegraphics*[]{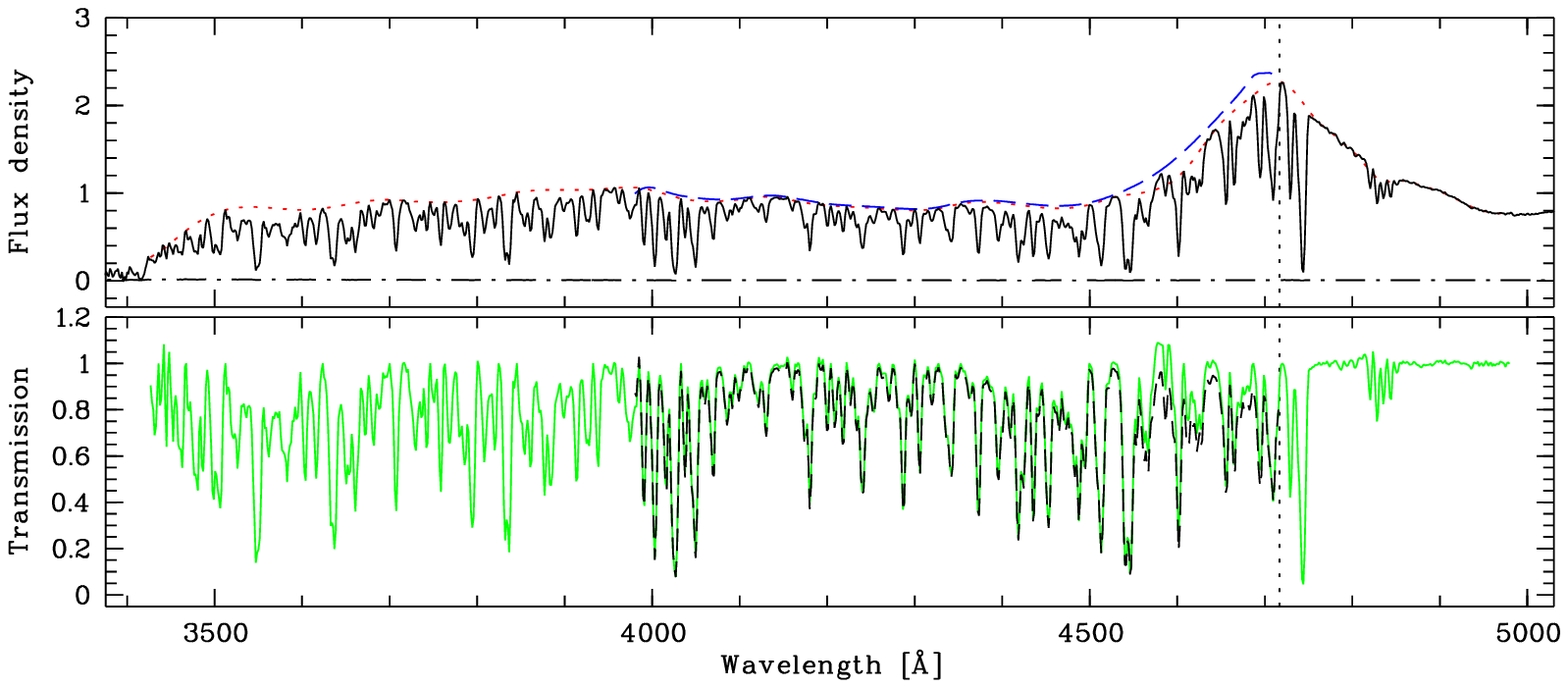}}
\caption{HE 2347$-$4342}
\end{figure}
\begin{figure}
\resizebox{\hsize}{!}{\includegraphics*[]{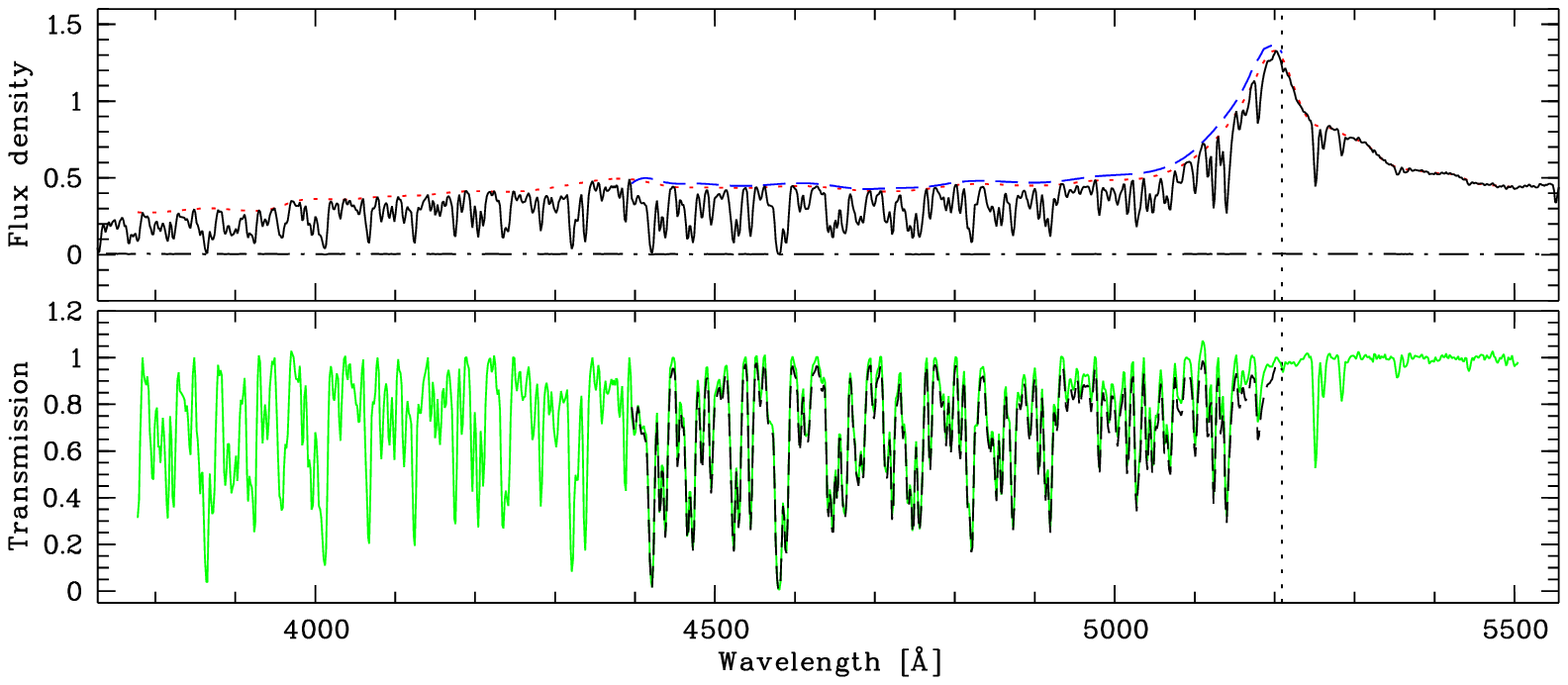}}
\caption{PKS 2126$-$15}
\end{figure}
\begin{figure}
\resizebox{\hsize}{!}{\includegraphics*[]{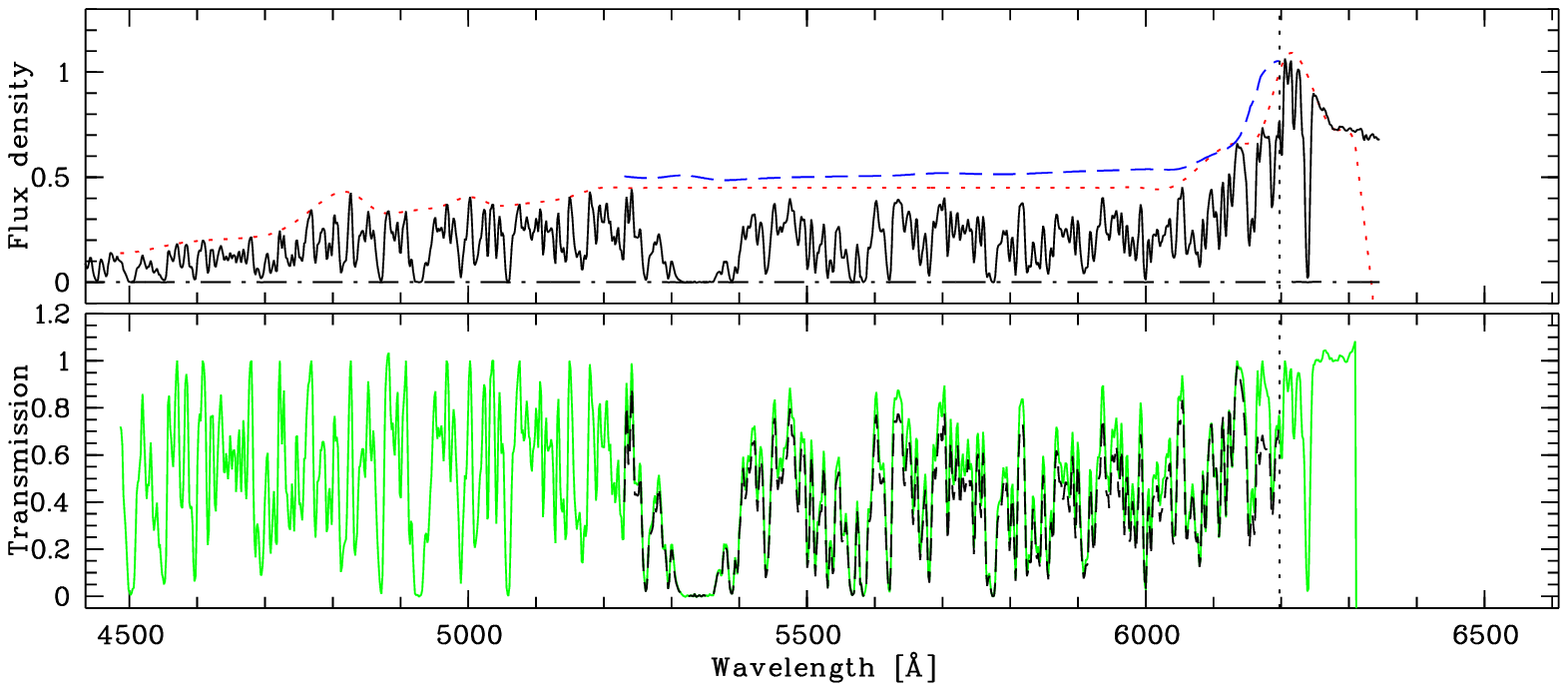}}
\caption{Q 0000$-$26}
\end{figure}
\begin{figure}
\resizebox{\hsize}{!}{\includegraphics*[]{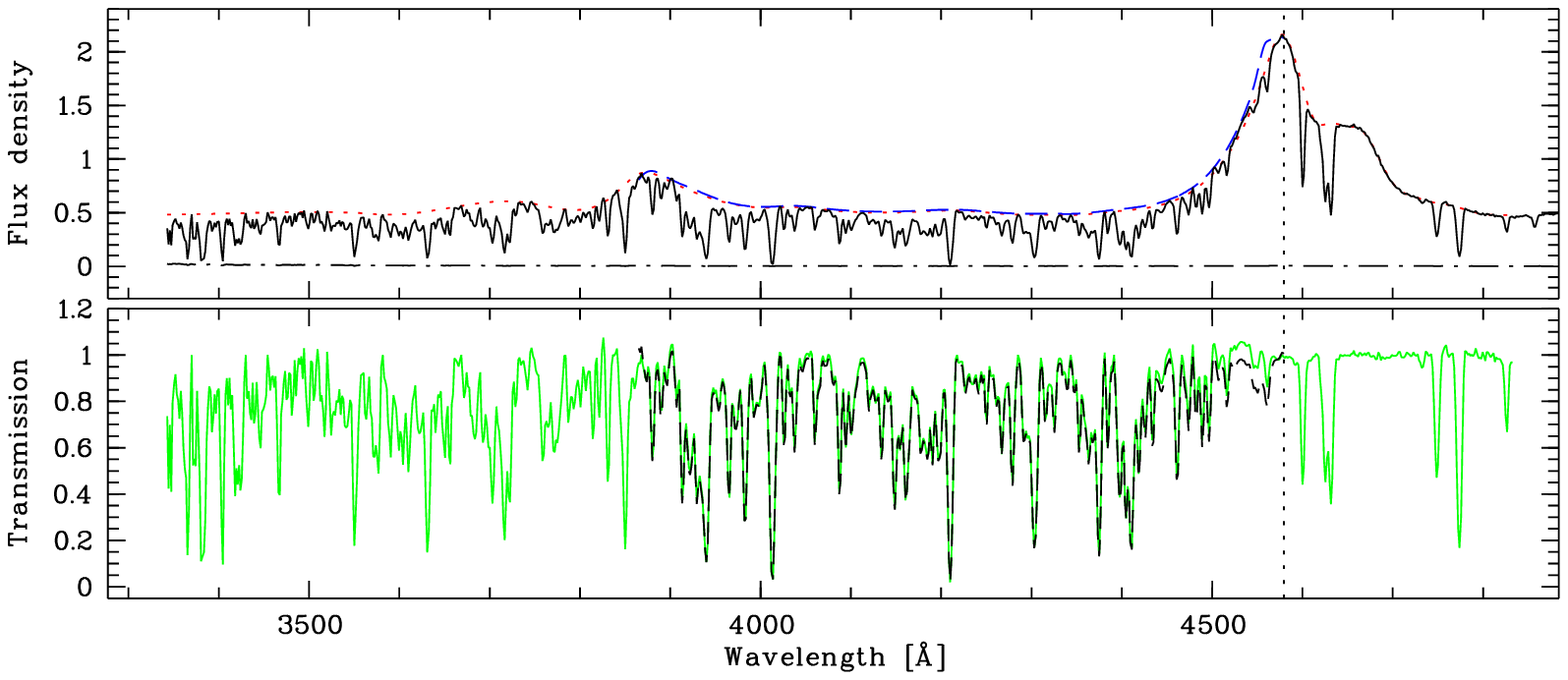}}
\caption{Q 0002$-$422}
\end{figure}
\begin{figure}
\resizebox{\hsize}{!}{\includegraphics*[]{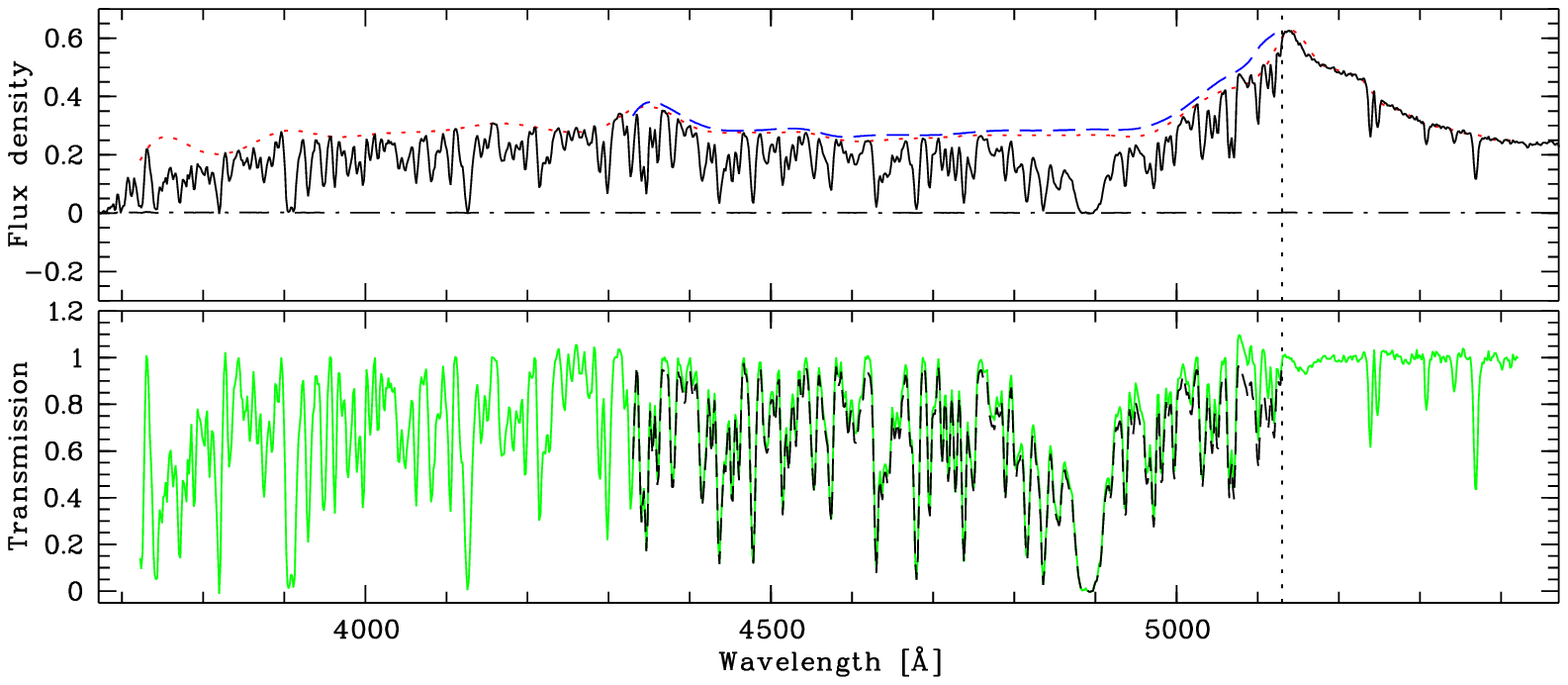}}
\caption{Q 0347$-$383}
\end{figure}
\begin{figure}
\resizebox{\hsize}{!}{\includegraphics*[]{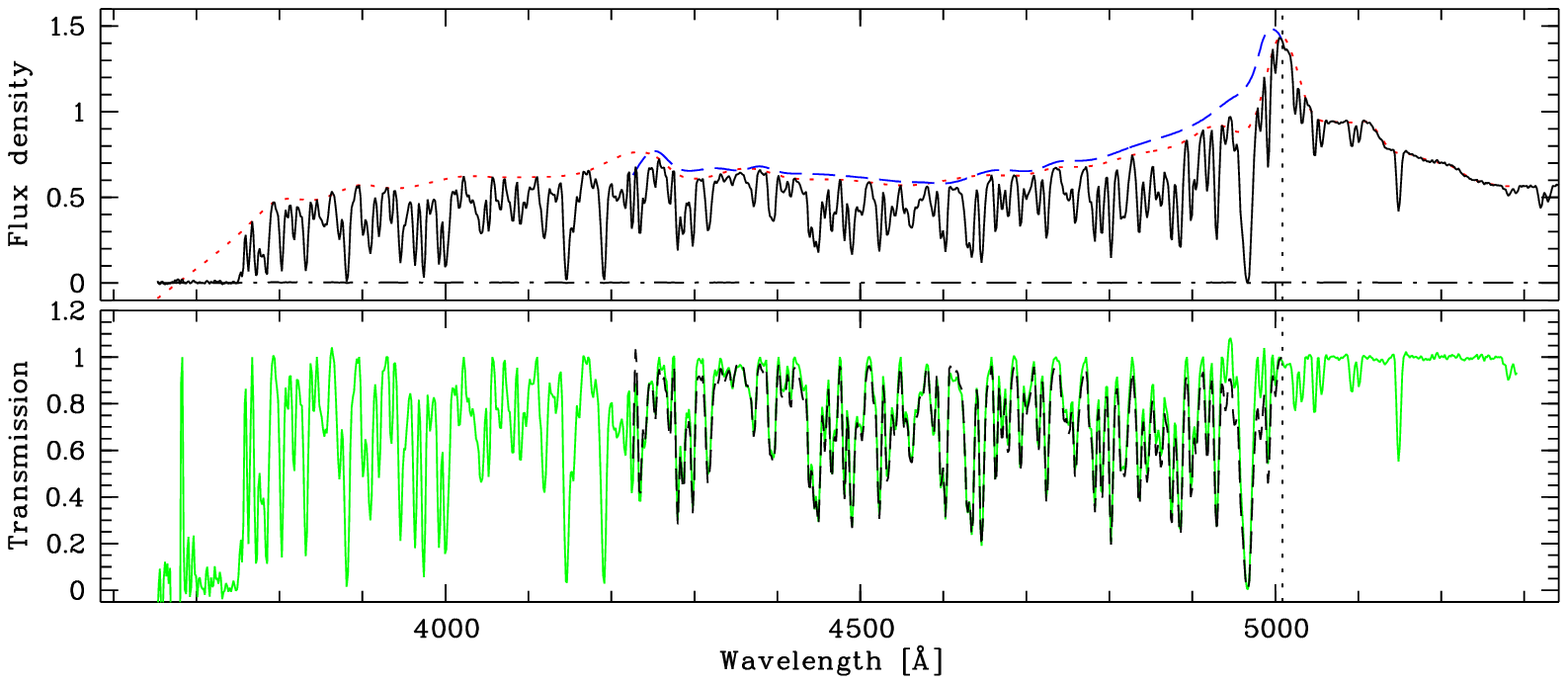}}
\caption{Q 0420$-$388}
\end{figure}
\begin{figure}
\resizebox{\hsize}{!}{\includegraphics*[]{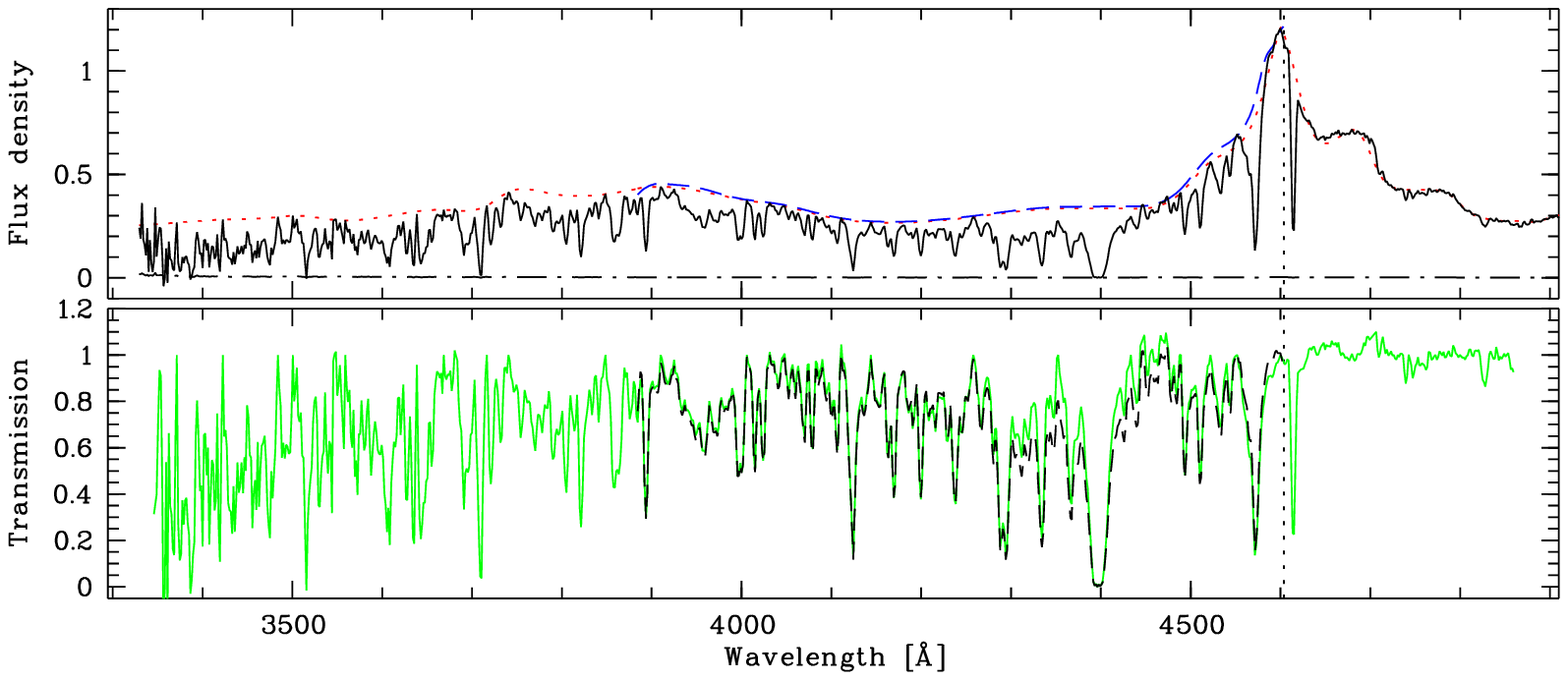}}
\caption{Q 0913$+$0715}
\end{figure}
\begin{figure}
\resizebox{\hsize}{!}{\includegraphics*[]{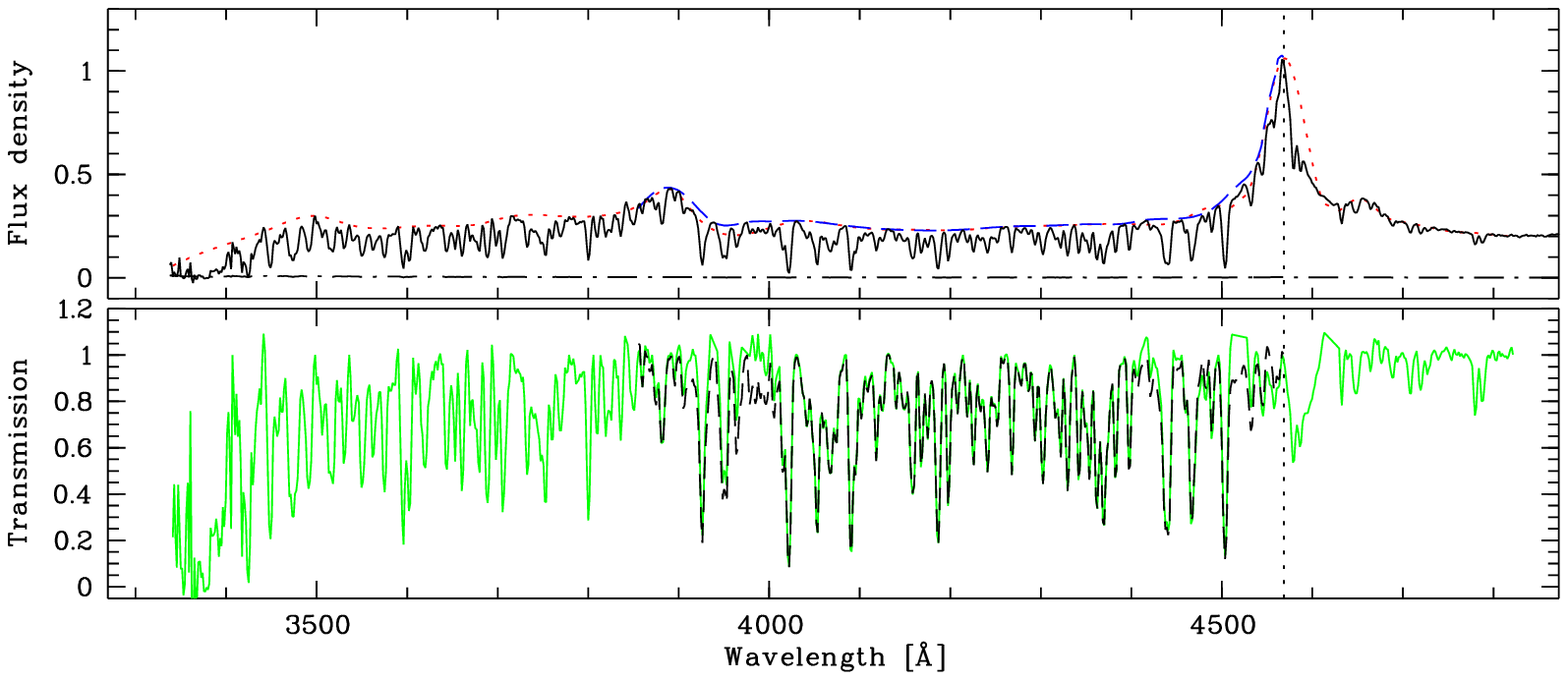}}
\caption{Q 1151$+$0651}
\end{figure}
\begin{figure}
\resizebox{\hsize}{!}{\includegraphics*[]{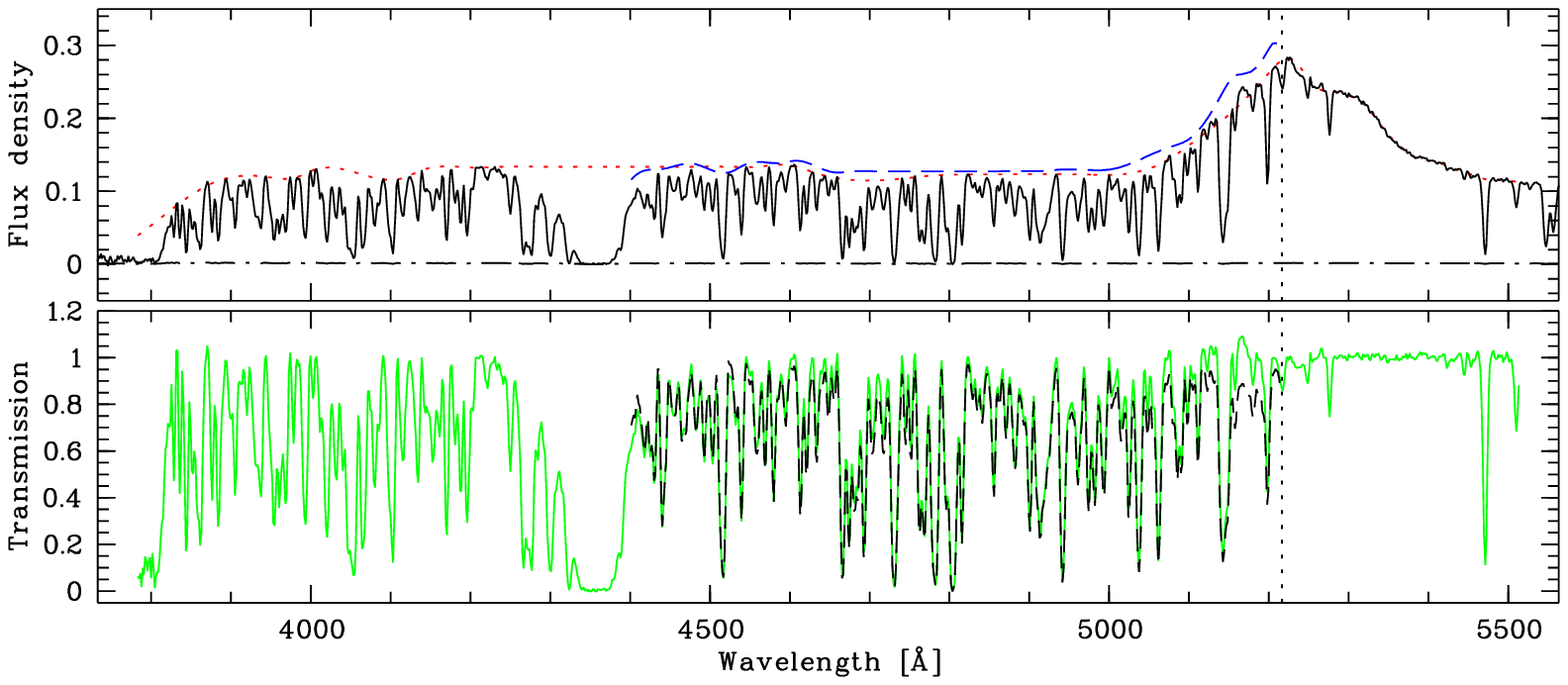}}
\caption{Q 1209$+$0919}
\end{figure}
\begin{figure}
\resizebox{\hsize}{!}{\includegraphics*[]{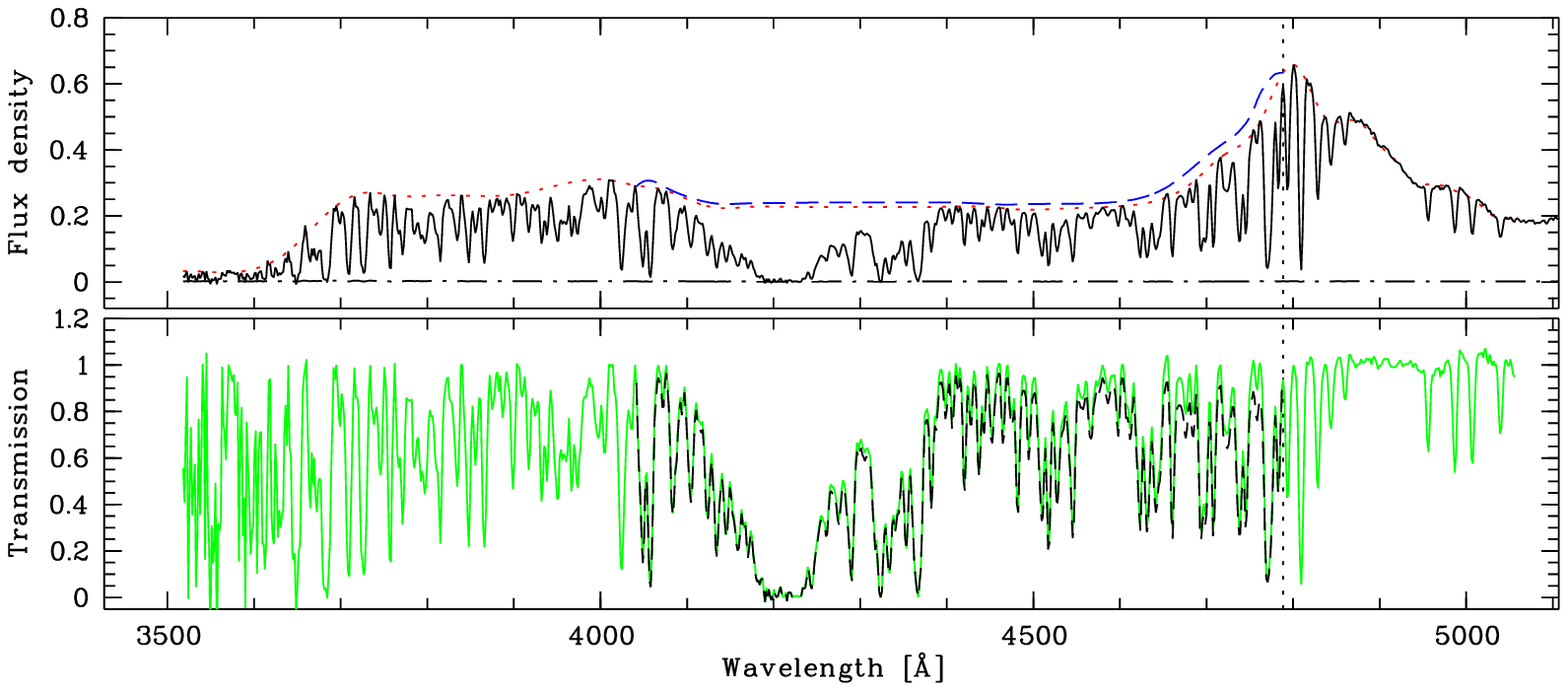}}
\caption{Q 1223$+$1753}
\end{figure}
\begin{figure}
\resizebox{\hsize}{!}{\includegraphics*[]{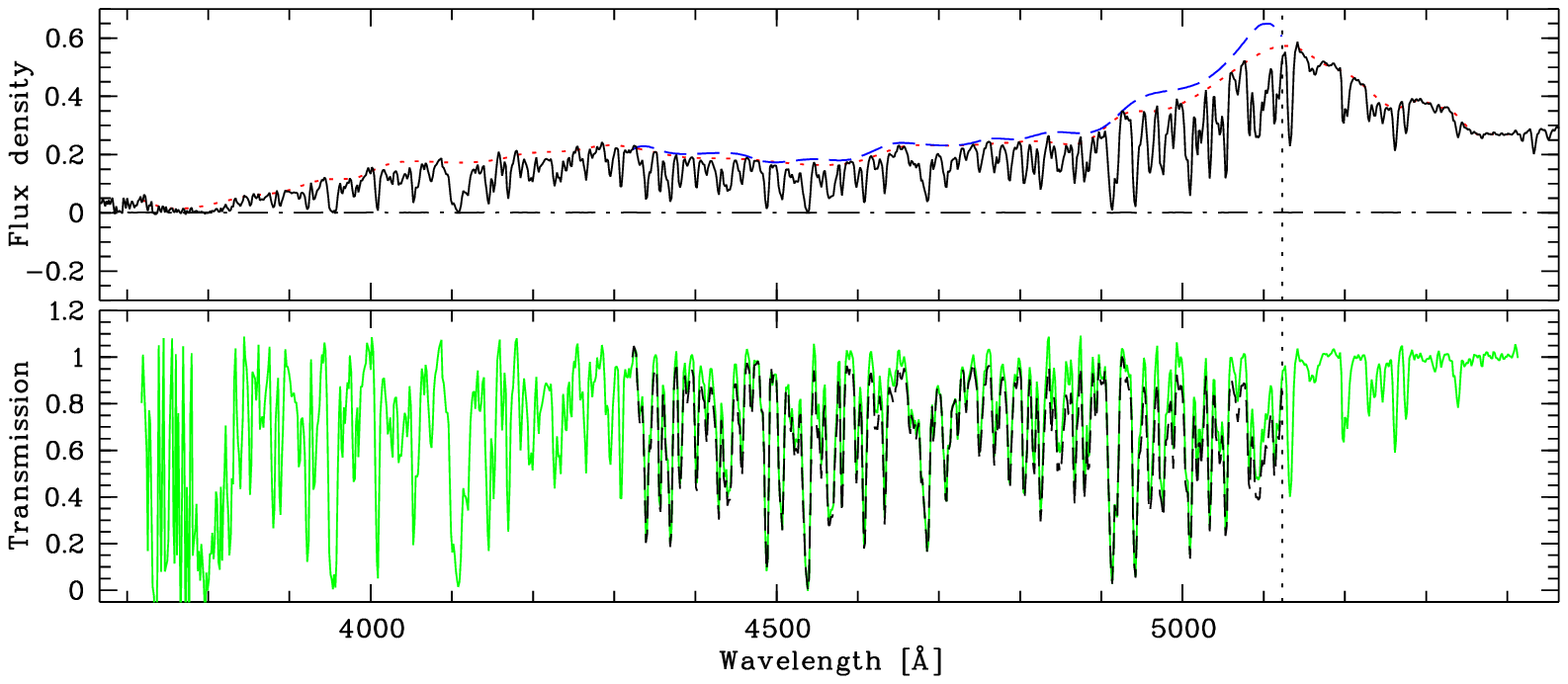}}
\caption{Q 2139$-$4434}
\end{figure}
\end{appendix}
\end{document}